\documentclass[aps,preprint,amsmath]{revtex4-1}

\usepackage{graphicx}
\usepackage{relsize}

\begin{document}


\title{
Spatiotemporal Structure of Aeolian Particle Transport on Flat Surface
}


\author{Hirofumi Niiya}
\author{Kouichi Nishimura}
\affiliation{Graduate School of Environmental Studies, Nagoya University, Nagoya, 464-8601, Japan}



\date{\today}

\begin{abstract}
We conduct numerical simulations based on a model of blowing snow to reveal the long-term properties and equilibrium state of aeolian particle transport from $10^{-5} \hspace{0.5 ex} \mathrm{m}$ to $10 \hspace{0.5 ex} \mathrm{m}$ above the flat surface.
The numerical results are as follows.
(i)
Time-series data of particle transport are divided into development, relaxation, and equilibrium phases,
which are formed by rapid wind response below $10 \hspace{0.5 ex} \mathrm{cm}$ and gradual wind response above $10 \hspace{0.5 ex} \mathrm{cm}$.
(ii)
The particle transport rate at equilibrium is expressed as a power function of friction velocity,
and the index of $2.35$ implies that most particles are transported by saltation.
(iii)
The friction velocity below $100 \hspace{0.5 ex} \mu\mathrm{m}$ remains roughly constant and lower than the fluid threshold at equilibrium.
(iv)
The mean particle speed above $300 \hspace{0.5 ex} \mu\mathrm{m}$ is less than the wind speed,
whereas that below $300 \hspace{0.5 ex} \mu\mathrm{m}$ exceeds the wind speed because of descending particles.
(v)
The particle diameter increases with height in the saltation layer,
and the relationship is expressed as a power function.
Through comparisons with the previously reported random-flight model,
we find a crucial problem that empirical splash functions cannot reproduce particle dynamics at a relatively high wind speed.
\end{abstract}


\maketitle

\section{Introduction}

Snow/sand erosion and deposition due to wind emit many deposited particles into the atmosphere,
and drifting snow and dust storms are generated as massive particles are transported.
Additionally, fluid-particle and particle-particle interactions increase spatial heterogeneity,
resulting in the formation of microscopic and macroscopic structures on snow/ice surfaces, sand deserts, and beaches.
For example, wind ripples and dunes are observed in natural fields.
The particle transport by the aeolian processes is a key factor to understand the morphodynamics of objects.
Generally, aeolian particle transport is maintained and developed through four physical sub-processes~\cite{bagnold1941physics}:
aerodynamical entrainment, wind-blown particle dynamics, splash caused by particle-granular bed collision, and wind modification
(Fig.~\ref{fig:1}).
In the equilibrium state, the wind profile is almost fixed because of the momentum exchange with particles,
whereas the dynamics of particles ejected from granular beds consists of three different modes:
creep, saltation, and suspension.
These motions strongly depend on wind speed and particle diameter;
thus, the spatial structure is complex in particle transport.

Recently, in order to measure the dynamics of each particle in transport,
snow/sand particle counter (SPC) and particle tracking velocimetry (PTV) have been applied to field observations and wind tunnel experiments~\cite{mann2000profile, nishimura2005blowing, nishimura2014snow, sugiura2000wind, yang2007height, gromke2014snow, walter2014experimental}.
SPC estimates the diameter and speed of each particle at an arbitrary point when the particle passes through that point,
whereas PTV directly calculates particle diameter and velocity from two-dimensional images.
In both systems, it is difficult to accurately measure the dynamics of particles near the surface because of the overlapped image of particles.
Lagrangian and turbulent diffusion theories, which are remarkable approaches,
have been used in the saltation and the suspension layer, respectively~\cite{shao1999numerical, nemoto2004numerical, kok2009comprehensive, creyssels2009saltating, huang20153, gauer2001numerical, kok2012physics}.
The former reproduces the detailed structure of local transport by computing the trajectory of each particle,
but the dynamics of particles are dependent on statistical functions (hereafter, splash function) characterizing the splash process.
The latter predicts global transport on actual landforms with the continuum approximation of blown particles,
although it is not suitable for transport with high particle inertia such as saltation.

In natural fields,
the granular bed consists of particles of various sizes,
and particles entrained from the bed exhibit different motions depending on the diameter.
The collision between a blown particle and the bed (i.e., the splash process) plays a key role in,
for example, the formation of saltation and suspension layers, entrainment of new particles, and particle velocity after collision.
Therefore, it is important to understand the particle transport property to calculate the dynamics of particles near surface.
In this study, we conduct numerical simulations based on the random-flight model~\cite{nemoto2004numerical} of blowing snow,
in which the splash function was measured in Sugiura et al.'s wind tunnel experiments~\cite{sugiura2000wind} using snow particles.
To reveal the spatiotemporal structure in aeolian particle transport including the saltation and suspension layers,
numerical simulations calculate the dynamics of each snow particle and wind speed profile from the vicinity of the surface ($10^{-5} \hspace{0.5 ex} \mathrm{m}$) to $10 \hspace{0.5 ex} \mathrm{m}$ above the surface.

\section{Model}

\begin{figure}[t]
\centering
\includegraphics[width=1.0\linewidth]{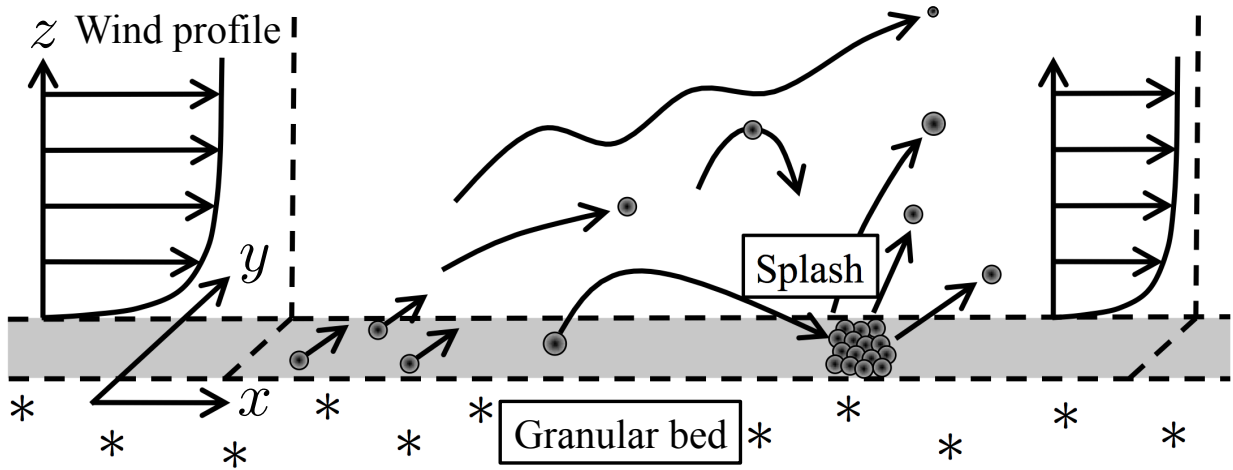}
\caption{
Schematic of four physical sub-processes in aeolian particle transport:
aerodynamical entrainment, wind-blown particle dynamics, splash, and wind modification.
}
\label{fig:1}
\end{figure}
%

This model simulates one-dimensional ($z$-axis) fluid dynamics and two-dimensional ($x$-$z$ plane) particle dynamics on the basis of the random-flight model~\cite{nemoto2004numerical}.
Here, the $x$-axis and $z$-axis represent the wind (horizontal) direction and the vertical direction, respectively
(Fig.~\ref{fig:1}).
As computational processes, four physical sub-processes in the aeolian particle transport are considered:
aerodynamical entrainment, wind-blown particle dynamics, splash process, and wind modification.
In this model, the fluid is treated as a turbulent boundary layer flow,
and particles with various sizes follow Newton's laws of motion.
The fluid-particle momentum exchange is expressed by the particle entrainment from the surface by wind and the air drag acting on each particle.
The following subsections explain the details of physical sub-processes.

\subsection{Wind modification}

The wind velocity $\mbox{\boldmath $u$}$ is simplified as $\mbox{\boldmath $u$} = (\overline{u}, w')$;
$\overline{u}$ is the horizontal component of time-averaged mean flow
and
$w'$ is the vertical component of turbulent fluctuation.
These two variables are calculated using the Reynolds-averaged Navier-Stokes equation and Lagrangian stochastic model, respectively.

By assuming uniform horizontal flow ($\partial / \partial x \approx 0$) and well-developed turbulent flow in the vertical direction,
the governing equation of $\overline{u}$ is expressed as:
\begin{eqnarray}
\rho_f \frac{\partial \overline{u}}{\partial t}
=
\frac{\partial \tau}{\partial z}
-
\sum_{i=1}^{N} \frac{F_d^i}{V_f},
\hspace{1 ex}
\label{eq:mean-flow}
\tau
=
\rho_f
\left(
\kappa z \frac{\partial \overline{u}}{\partial z}
\right)^2,
\end{eqnarray}
where $\rho_f$ is the fluid density;
$\tau$ and $\kappa$ are the fluid shear stress and Karman constant (0.4), respectively;
$V_f$ is the volume of the fluid computational mesh;
and $N$ and $F_d^i$ are the number of particles and horizontal air drag acting on the $i$th particle within $V_f$, respectively.
Here, $\tau$ is described by Prandtl's mixing length theory (i.e., the simplest turbulence model),
which ignores the viscous stress due to the well-developed turbulent flow.

The vertical turbulent fluctuation $w'$ is calculated for each particle because of the spatial decorrelation.
In the case of turbulence with spatial uniformity and isotropy,
the time variation of $w'_i$ acting on the $i$th particle is generally represented using Euler statistics and the Kolmogorov similarity law~\cite{wilson1996review}:
\begin{eqnarray}
w'_i(t+\Delta t)
=
\left(
1 - \frac{\Delta t}{T_L^*}
\right)
w'_i(t)
+
\sigma_w \sqrt{\frac{2 \Delta t}{T_L^*}} \eta (t),
\label{eq:turbulent-fluctuation}
\end{eqnarray}
where $\Delta t$ is the time step,
$T_L^*$ is the Lagrangian time scale with considering the particle inertia,
$\sigma_w$ is the turbulent intensity,
and $\eta (t)$ is a random number generated by the standard normal distribution $N(0,1)$. 
Using empirical formulae in the boundary layer of a neutral atmosphere~\cite{hunt1979lagrangian, hunt1985saltating},
two unknown parameters ($T_L^*, \sigma_w$) are given as functions of friction velocity $u^* = \sqrt{\tau / \rho_f}$:
\begin{eqnarray}
T_L^*
=
\frac{T_L}
{1 + A \left( V_R / \sigma_w \right)^{2/3} \left( T_L / \Delta t \right)^{1/3} },
\hspace{1 ex}
\sigma_w
=
1.3 u^*,
\label{eq:time-intensity}
\end{eqnarray}
where $A$ is a constant (0.5),
$V_R$ is the relative speed between the wind and $i$th particle expressed as $V_R = |\mbox{\boldmath $u$} - \mbox{\boldmath $v$}_i|$,
$T_L$ is the Lagrangian time scale ignoring the particle inertia defined as $T_L = z/(2\sigma_w)$, respectively.

\subsection{Aerodynamical entrainment}
\label{sec:2-2}

Deposited particles start to migrate if the wall friction velocity $u_w^*$ acting on the surface exceeds the fluid threshold $u_f$.
For the snow particle,
$u_f$ and the diameter have various values owing to the cohesion of particles,
although we use a constant value as $u_f$ in this model.
Then, the number of entrained particles $N_e$ per unit time and unit area is expressed based on experiments with monodisperse particles as~\cite{shao1999numerical}:
\begin{eqnarray}
N_e
=
\xi u_w^*
\left[
1 - \left( \frac{u_f}{u_w^*} \right)^2
\right]
\overline{d}^{-3},
\hspace{1 ex}
\xi
=
\frac{6 \rho_f}{a \pi \rho_p},
\label{eq:number-particles}
\end{eqnarray}
where $\xi$ is a dimensionless parameter equal to the ratio between fluid and particle density $\rho_p$;
$a$ and $\pi$ are a constant (0.5) and the circumference ratio, respectively;
and $\overline{d}$ is the mean particle diameter in the granular bed.

The diameter of an entrained particle is selected from the particle size distribution in the granular bed,
and the initial coordinate is randomly set on the $x$-$y$ plane to contact the surface ($z=d/2$).
In this model,
the particle does not move in the $y$ direction;
thus, the initial velocity is given as
\begin{eqnarray}
v_x
=
a u_w^*,
\hspace{1 ex}
v_z
=
\sqrt{2 g d},
\label{eq:initial-velocity}
\end{eqnarray}
where $a$ has the same value as in eq.~(\ref{eq:number-particles}) and $g$ is the acceleration due to gravity.
Here,
the specific form of $v_x$ is defined by the hypothesis of eq.~(\ref{eq:number-particles}),
whereas the specific form of $v_z$ is set to reach the particle diameter at most.

\subsection{Wind-blown particle dynamics}

Particles ejected from the granular bed are assumed to be the irrotational hard-spherical grains.
We also ignore the collision between particles by considering a low particle number density;
therefore, gravity and air drag are taken into account as forces acting on each particle.
According to the above assumptions,
the $i$th particle dynamics in saltation and suspension is expressed as
\begin{eqnarray}
\frac{d \mbox{\boldmath $x$}_i}{dt}
=
\mbox{\boldmath $v$}_i,
\hspace{1 ex}
m_i \frac{d \mbox{\boldmath $v$}_i}{dt}
=
- \mbox{\boldmath $e$}_z m_i g
+ C_d \frac{\rho_f |\mbox{\boldmath $u$} - \mbox{\boldmath $v$}_i| (\mbox{\boldmath $u$} - \mbox{\boldmath $v$}_i)}{2} S_i,
\label{eq:particle-dynamics}
\end{eqnarray}
where $m_i$, $\mbox{\boldmath $x$}_i$, $\mbox{\boldmath $v$}_i$, and $S_i$ are the mass, coordinate, velocity, and cross-sections of the $i$th particle, respectively;
$\mbox{\boldmath $e$}_z$ is the unit vector parallel to the $z$-axis;
and $C_d$ is the drag coefficient defined by a function of the particle Reynolds number $Re_p$:
\begin{eqnarray}
C_d
=
\frac{24}{Re_p} + \frac{6}{1 + Re_p^{1/2}} + 0.4.
\label{eq:drag-coefficient}
\end{eqnarray}
Equation~(\ref{eq:drag-coefficient}) is the approximate formula for a single spherical particle~\cite{morsi1972investigation},
and it is roughly applicable for a high particle Reynolds number:
$Re_p \approx 10^5$.

\subsection{Splash process}
\label{sec:2-4}

Splash occurs if the $i$th particle collides with the surface;
that is, the particle height $z_i$ is less than the half of the diameter ($d/2$) and the vertical speed $v_{iz}$ is negative.
The splash process in this model is represented by empirical statistical functions (splash functions),
which are obtained from wind tunnel experiments to detect each particle-bed collision in the snow particle transport~\cite{sugiura2000wind}.

\subsubsection{Splash functions}
\label{sec:2-4-1}

\begin{figure}[t]
\centering
\includegraphics[width=0.6\linewidth]{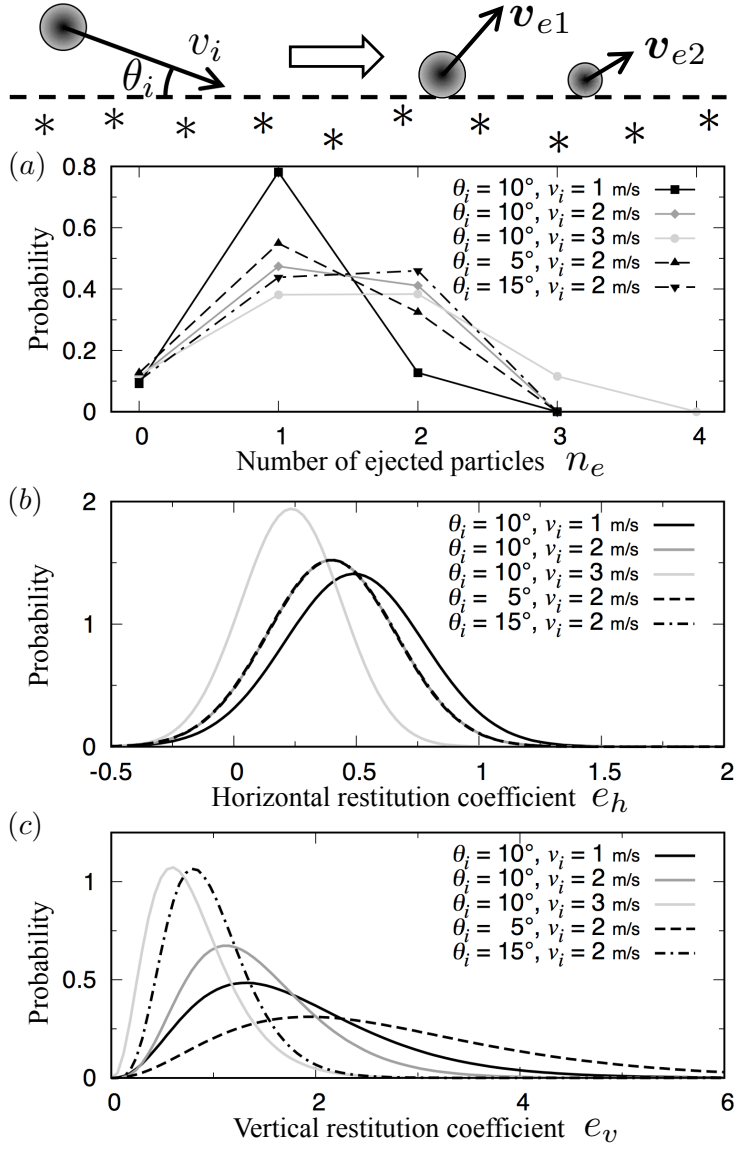}
\caption{
Probability density functions in splash process calculated from single incident particle with angle $\theta_i$ and speed $v_i$:
(a) number of ejected particles $n_e$,
(b) horizontal restitution coefficient $e_h$,
and (c) vertical restitution coefficient $e_v$.
Gray-scaled and dashed lines indicate incident speed and angle dependency, respectively.
Note that functions of $e_h$ at $v_i =$ 2 m s$^{-1}$ overlap with each other.
}
\label{fig:2}
\end{figure}

We directly apply the splash functions proposed by Sugiura et al.~\cite{sugiura2000wind} to the rebounded and splashed particles,
but in the case of number of particles,
the splash function is modified to be a smooth function for change in input parameters.
Splash functions estimate three values for clarifying the dynamics of particles by utilizing the incident angle $\theta_i$ and speed $v_i$:
number of ejected particles $n_e$ including the rebounded particles,
and horizontal and vertical restitution coefficients $(e_h, e_v) \equiv (v_{ex}/v_{ix}, v_{ez}/|v_{iz}|)$ with the ejected particle velocity $\mbox{\boldmath $v$}_e$ and incident particle velocity $\mbox{\boldmath $v$}_i$
(Fig.~\ref{fig:2}).
According to the experimental results obtained by Sugiura et al.~\cite{sugiura2000wind},
the distributions of $n_e$, $e_h$, and $e_v$ are fitted by binomial, normal, and gamma distributions, respectively, as follows:
\begin{eqnarray}
n_e
\in
B(m,p),
\hspace{1 ex}
e_h
\in
N(\mu,\sigma^2),
\hspace{1 ex}
e_v
\in
\Gamma(\alpha,\beta),
\label{eq:splash-functions}
\end{eqnarray}
where $m$, $p$, $\mu$, $\sigma^2$, $\alpha$, and $\beta$ are parameters characterizing each distribution.
These parameters are also expressed as functions of $\theta_i$ and $v_i$~\cite{sugiura2000wind}:
\begin{eqnarray}
m
&=&
\frac{0.64 \theta_i^{0.22} v_i^{0.62}}{0.8 \theta_i^{0.11}v_i^{0.31} - 0.05 \theta_i^{0.36} v_i^{1.58}},
\label{eq:splash-number1}\\
p
&=&
1 - 0.06 \theta_i^{0.25} v_i^{1.27},
\label{eq:splash-number2}
\end{eqnarray}
\begin{eqnarray}
\mu
=
\left\{
\begin{array}{ll}
0.48 \theta_i^{0.01} & v_i \in (0,1.27],\\
0.48 \theta_i^{0.01} \left( \dfrac{v_i}{1.27} \right)^{-\log(\frac{v_i}{1.27})} & v_i \in (1.27,+\infty],
\end{array}
\right.
\label{eq:splash-horizontal1}\\
\sigma^2
=
\left\{
\begin{array}{ll}
0.08 \theta_i^{0.01} & v_i \in (0,1.34],\\
0.08 \theta_i^{0.01} \left( \dfrac{v_i}{1.34} \right)^{-\log(\frac{v_i}{1.34})} & v_i \in (1.34,+\infty],
\end{array}
\right.
\label{eq:splash-horizontal2}
\end{eqnarray}
\begin{eqnarray}
\alpha
=
\left\{
\begin{array}{ll}
1.22 \theta_i^{0.47} & v_i \in (0,0.84],\\
1.22 \theta_i^{0.47} \left( \dfrac{v_i}{0.84} \right)^{\log(\frac{v_i}{0.84})} & v_i \in (0.84,1.23],\\
1.22 \theta_i^{0.47} \left( \dfrac{v_i}{0.84} \right)^{\log(\frac{v_i}{0.84})} \left( \dfrac{v_i}{1.23} \right)^{-2\log(\frac{v_i}{1.23})} & v_i \in (1.23,+\infty],
\end{array}
\right.
\label{eq:splash-vertical1}\\
\beta
=
\left\{
\begin{array}{ll}
12.85 \theta_i^{-1.41} & v_i \in (0,0.84],\\
12.85 \theta_i^{-1.41} \left( \dfrac{v_i}{0.84} \right)^{-\log(\frac{v_i}{0.84})} & v_i \in (0.84,1.23],\\
12.85 \theta_i^{-1.41} \left( \dfrac{v_i}{0.84} \right)^{-\log(\frac{v_i}{0.84})} \left( \dfrac{v_i}{1.23} \right)^{\log(\frac{v_i}{1.23})} & v_i \in (1.23,+\infty],
\end{array}
\right.
\label{eq:splash-vertical2}
\end{eqnarray}
where the physical units of $\theta_i$ and $v_i$ are degree in the range of $0^{\circ}$ to $90^{\circ}$ and $\mathrm{m \hspace{0.5 ex} s^{-1}}$, respectively.

Here, $n_e =$~0 and 1 indicate the deposition and rebound of an incident particle, respectively,
whereas $n_e \ge 2$ indicates the emission of splashed particles.
In the case of $n_e \ge 2$,
the diameter of each splashed particle $d$ is selected from the particle size distribution of the granular bed.
The horizontal position of them is set to be the same as that of an incident particle,
whereas the vertical position is given as $z = d/2$.
The ejected velocities are calculated using $e_h$ and $e_v$ as $(v_{ex}, v_{ez}) = (e_h v_{ix}, e_v |v_{iz}|)$.
Note that if the total ejected kinematic energy exceeds the incident kinematic energy,
we recalculate the splash process.

\subsubsection{Effect of incident angle and speed on functions}
\label{sec:2-4-2}

Number of ejected particles:
$n_e$ increases with increase in $\theta_i$ and $v_i$,
which increase the incident speed perpendicular to the surface: $v_{iz} = - v_i \sin \theta_i$.
The emission of deposited particles is enhanced by the increase in $v_{iz}$
(Fig.~\ref{fig:2}(a)).
%

Horizontal restitution coefficient:
$e_h$ decreases with increase in $v_i$,
but it is nearly independent of $\theta_i$
(Fig.~\ref{fig:2}(b)).
In the splash process, momentum is transmitted from an incident particle to deposited particles,
and the increase in $v_i$ increases the efficiency of the momentum exchange because of the increase in $n_e$.
Then, the momentum (i.e., velocity) of the rebounded particle does not increase drastically with increase in $v_i$,
which leads to the decrease in $e_h$.
On the other hand, the increase in $\theta_i$ slightly increases $n_e$,
although the change in the horizontal incident speed $v_{ix} = v_i \cos \theta_i$ is quite small at $\theta_i = 5^{\circ}, 10^{\circ}, 15^{\circ}$,
which were the values used in the data set of experiments by Sugiura et al.~\cite{sugiura2000wind}.
Because of the low dependence of $n_e$ and $v_{ix}$ on $\theta_i$,
they seem to have distributions similar to that of $e_h$.
%

Vertical restitution coefficient:
$e_v$ decreases with increase in $\theta_i$ and $v_i$
(Fig.~\ref{fig:2}(c)).
The change in $e_v$ with $v_i$ is explained by the momentum exchange from the incident particle to deposited particles.
In the case of the low $\theta_i$ or the low $v_i$ (i.e., the small $n_e$),
the momentum in the ascent direction is mainly transmitted to the incident particle.
Whereas in the case of the larger $\theta_i$ or the higher $v_i$ (i.e., the relatively large $n_e$),
the momentum in the ascent direction is also utilized for entrained particles.
This difference moderates the increase in vertical ejected speed $v_{ez}$,
causing the decrease in $e_v$.


\subsection{Setup of numerical simulations}
\label{sec:2-5}

Numerical simulations of the model are conducted on a flat surface with a constant roughness length $z_0 = 10^{-5}$~m,
which means the surface asperity and is fixed during the simulation.
The domain is a cuboid with dimensions of $L = 2 \hspace{0.5 ex} \mathrm{cm}$, $W = 1 \hspace{0.5 ex} \mathrm{cm}$, and $H = 10 \hspace{0.5 ex} \mathrm{m}$ height.
Although the calculation is two-dimensional, $W$ is used for the aerodynamical entrainment (Sect.~\ref{sec:2-2}).
The fluid mesh is logarithmically generated.

As the vertical boundary condition for the fluid,
the mean horizontal wind velocity $\overline{u}$ is given as zero below the height of $z_0$,
and the friction velocity $u^*$ is given as constant at the top:
$\overline{u}(z \le z_0) = 0$ and $u^*(H) = \mathrm{constant}$.
In this assumption, the wind velocity at the top can vary with time.
Additionally, the initial vertical turbulent fluctuation $w'_i$ acting on the $i$th particle is set as zero
when the particle is entrained by wind or splash.
The horizontal boundary condition for particles is set as periodic.
If the $i$th particle exceeds the top, i.e., $z_i > H$,
$\overline{u}$ at the particle coordinates is estimated by extrapolating the wind profile of $u^*(H)$.
Next, the initial condition is assumed to be steady wind of eq.~(\ref{eq:mean-flow}) without particles:
\begin{eqnarray}
\overline{u}(t=0, z)=
\left\{
\begin{array}{ll}
0 & z \le z_0 \\
\dfrac{u^*(H)}{\kappa} \log \left( \dfrac{z}{z_0} \right) & z > z_0
\end{array}
\right.
,
\hspace{1 ex}
N = 0.
\label{eq:initial-condition}
\end{eqnarray}
Here, $u^*(H)$ characterizes the wind intensity in numerical simulations;
therefore, we vary $u^*(H)$, which takes the values of $0.25, 0.3, 0.4, 0.5, 0.6 \hspace{0.5 ex} \mathrm{m \hspace{0.5 ex} s^{-1}}$.

For other parameters, the value for the air and the dry snow particle are used in the simulation.
Fluid and particle densities are fixed as $\rho_f = 1.2 \hspace{0.5 ex} \mathrm{kg \hspace{0.5 ex} m^{-3}}$ and $\rho_p = 900 \hspace{0.5 ex} \mathrm{kg \hspace{0.5 ex} m^{-3}}$, respectively,
and thus the dimensionless parameter $\xi$ in eq.~(\ref{eq:number-particles}) is roughly given as $10^{-3}$.
According to experiments by Sugiura et al.~\cite{sugiura2000wind},
the fluid threshold $u_f$ of compact snow particles is in the range of $0.19 \sim 0.25 \hspace{0.5 ex} \mathrm{m \hspace{0.5 ex} s^{-1}}$;
therefore, we assume $u_f = 0.20 \hspace{0.5 ex} \mathrm{m \hspace{0.5 ex} s^{-1}}$.
The particle size distribution at the granular bed is approximated by various functions depending on fields and experimental conditions,
although the gamma distribution is used in our simulations on the basis of the experiment by Gromke et al.~\cite{gromke2014snow} and the observation by Schmidt~\cite{schmidt1982vertical}.
The diameter $d$ of the entrained particle is selected from $\Gamma (3, 100)$,
where the mean, standard deviation, and peak are $300 \hspace{0.5 ex} \mu \mathrm{m}$, $100 \sqrt{3} \hspace{0.5 ex} \mu \mathrm{m}$, and $200 \hspace{0.5 ex} \mu \mathrm{m}$, respectively.
Note that $d$ is limited to within the range of $10 \hspace{0.5 ex} \mu \mathrm{m} \sim 1 \hspace{0.5 ex} \mathrm{mm}$.

\section{Results}

\begin{figure}[t]
\centering
\includegraphics[width=1.0\linewidth]{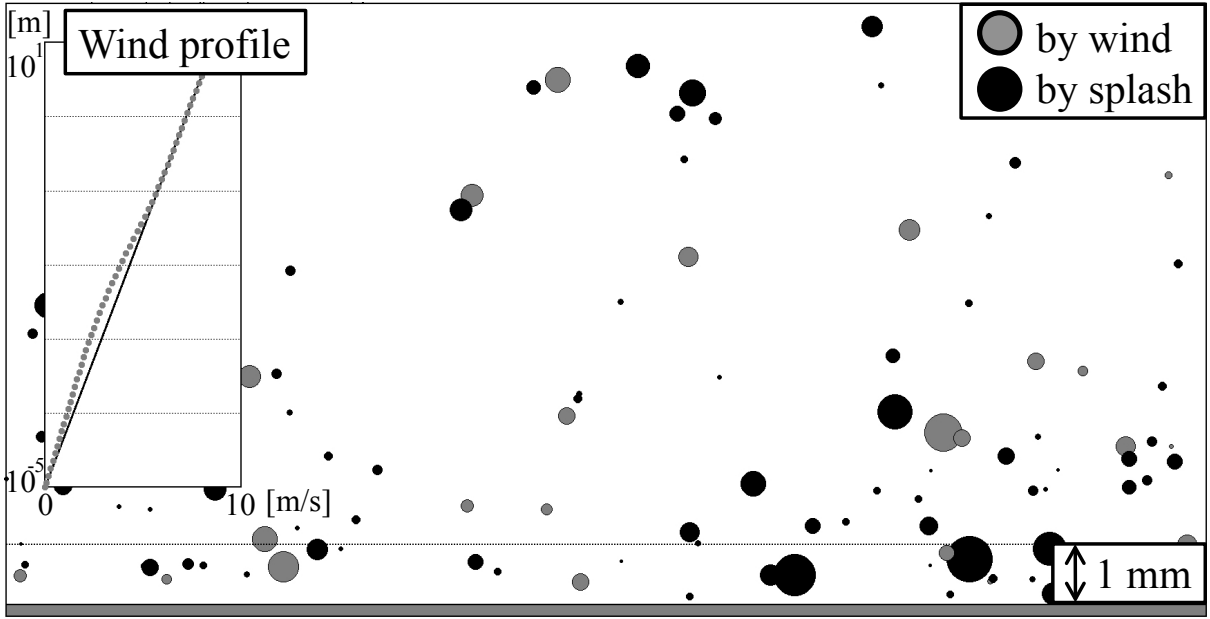}
\caption{
Simulation image at $t = 0.5 \hspace{0.5 ex} \mathrm{s}$ and $u^*(H) = 0.25 \hspace{0.5 ex} \mathrm{m \hspace{0.5 ex} s^{-1}}$.
The gray bottom rectangle denotes the granular bed, where new particles are emitted into the atmosphere.
Gray and black circles are particles ejected by aerodynamical entrainment and the splash process, respectively.
In the wind profile, the black solid line indicates the initial wind velocity given by eq.~(\ref{eq:initial-condition}),
whereas the gray dot indicates the calculated value at each fluid mesh.
}
\label{fig:3}
\end{figure}
%

In order to elucidate the developmental process and equilibrium state in aeolian particle transport,
we conduct numerical simulations of this model with different wind strength conditions.
Firstly, the long-term change in the particle transport is investigated under a weak wind condition consistent with the experimental conditions of Sugiura et al.~\cite{sugiura2000wind}.
Next, the wind-strength dependence of the particle transport is shown at the equilibrium state;
in particular, we focus on the spatial structure change and relationship between wind speed and particle speed.
Finally, the particle dynamics depending on the diameter is revealed.

\subsection{Temporal change in particle transport}
\label{sec:3-1}

\begin{figure}[t]
\centering
\includegraphics[width=1.0\linewidth]{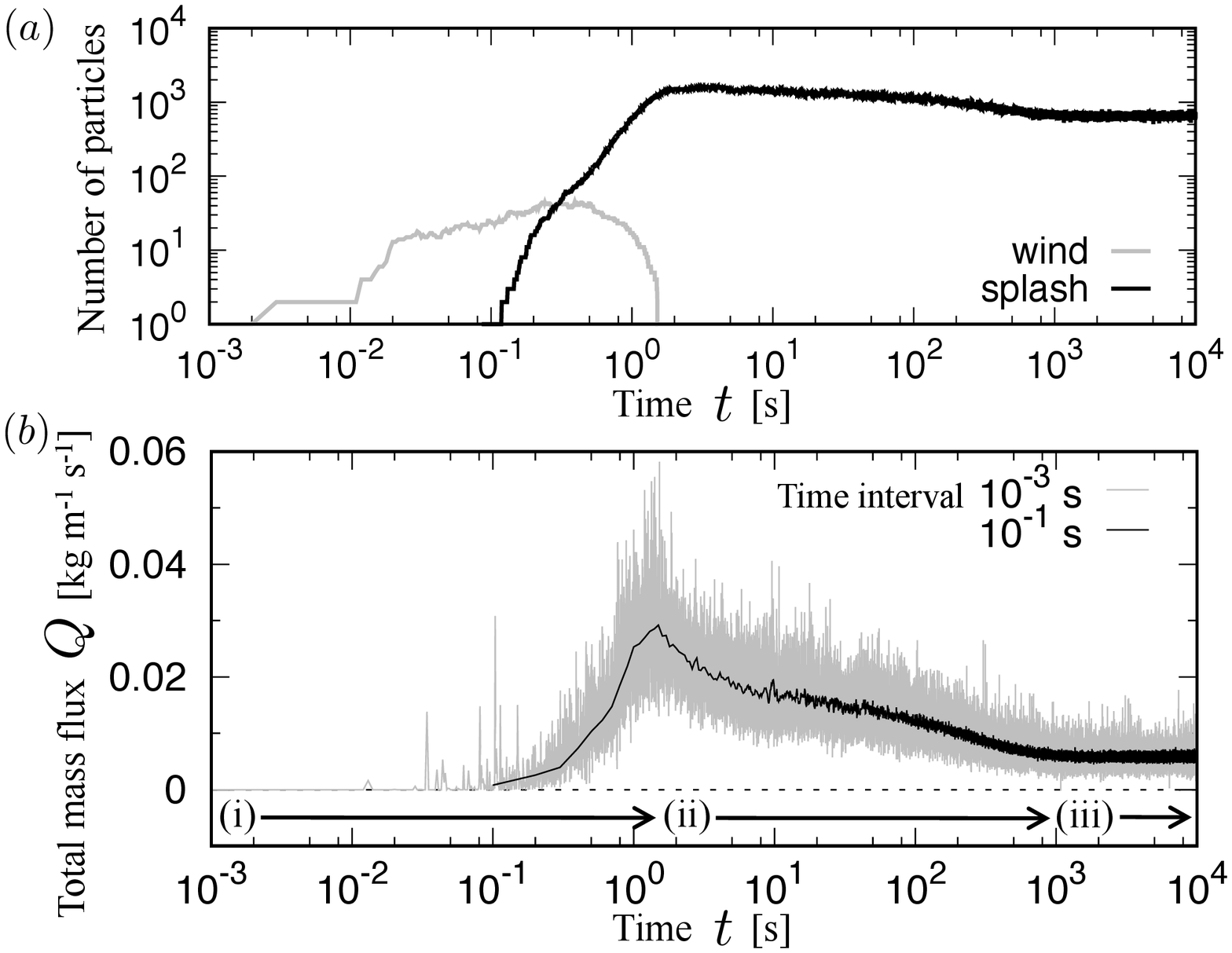}
\caption{
Time-series data at top friction velocity $u^*(H) = 0.25 \hspace{0.5 ex} \mathrm{m \hspace{0.5 ex} s^{-1}}$:
(a) number of particles entrained by wind (gray) and splash (black) and (b) total mass flux $Q$.
(b) The time variation of $Q$ is divided into three phases: (i) development, (ii) relaxation, and (iii) equilibrium.
Line colors show different time intervals to measure the mass flux:
$10^{-3} \hspace{0.5 ex} \mathrm{s}$ (gray) and $10^{-1} \hspace{0.5 ex} \mathrm{s}$ (black).
}
\label{fig:4}
\end{figure}
%

The numerical simulation is carried out at the top friction velocity $u^* (H) = 0.25 \hspace{0.5 ex} \mathrm{m \hspace{0.5 ex} s^{-1}}$,
in which the splash functions were those reported by Sugiura et al.~\cite{sugiura2000wind}.
Since the wall friction velocity $u_w^* (\equiv u^*(z_0))$ acting on the surface is greater than the fluid threshold $u_f = 0.2 \hspace{0.5 ex} \mathrm{m \hspace{0.5 ex} s^{-1}}$,
the aerodynamical entrainment of particles initially increases.
Figure~\ref{fig:3} shows the particle transport at $t = 0.5 \hspace{0.5 ex} \mathrm{s}$ in the vicinity of the surface.
Particles of various sizes are concurrently blown in the atmosphere, and they originate from wind and splash,
as shown by the gray and black circles in Fig.~\ref{fig:3}.
In this case, the mean horizontal wind speed $\overline{u}$ below $10 \hspace{0.5 ex} \mathrm{cm}$ decreases with the momentum exchange from the wind to particles.

\subsubsection{Number of particles and transport rate}
\label{sec:3-1-1}

Figure~\ref{fig:4}(a) shows time-series data for the numbers of particles blown by wind and splash,
which are denoted by gray and black lines.
The initial particle transport consists of particles entrained by only wind,
and the number of particles blown by wind increases with time.
Splashed particles occur from $t \approx 0.1 \hspace{0.5 ex} \mathrm{s}$, and their number increases drastically through the chain process of splash,
whereas the number of particles due to the aerodynamical entrainment immediately decreases and then disappears from the particle transport.
Therefore, the main particle entrainment shifts from the wind to the splash at the early stage ($t < 1 \hspace{0.5 ex} \mathrm{s}$).
After the disappearance of particles due to the wind,
the number of splashed particles gradually decreases and eventually remains at the same level after $t = 10^{3} \hspace{0.5 ex} \mathrm{s}$.

As the particle transport rate,
we define $Q$ as the integrated value of horizontal particle mass flux $q(z)$ from zero to infinity,
but the integral range in our simulations is given as $[0, H]$:
\begin{eqnarray}
Q
=
\int_{0}^{H} q(z) dz.
\label{eq:total-mass-flux}
\end{eqnarray}
Hereafter, $Q$ is called the total mass flux.
Figure~\ref{fig:4}(b) shows the time-series data of $Q$, which roughly reflects the number of particles shown in Fig.~\ref{fig:4}(a).
Gray and black lines denote the difference in measurement time interval for $Q$: $10^{-3} \hspace{0.5 ex} \mathrm{s}$ and $10^{-1} \hspace{0.5 ex} \mathrm{s}$.
According to Fig.~\ref{fig:4}(b),
we can categorize the state of aeolian transport into three phases:
(i) development ($t < 1 \hspace{0.5 ex} \mathrm{s}$),
(ii) relaxation ($t < 10^{3} \hspace{0.5 ex} \mathrm{s}$),
and (iii) equilibrium ($t \ge 10^{3} \hspace{0.5 ex} \mathrm{s}$).
$Q$ is a simple indicator to characterize the transport state,
but it is not always appropriate for understanding the spatial structure.
Thus, we check the friction velocity $u^*$, mean horizontal wind speed $\overline{u}$, and particle height in order to reveal the details of each phase.

\subsubsection{Structure transition of particle transport}
\label{sec:3-1-2}

\begin{figure}[t]
\centering
\includegraphics[width=1.0\linewidth]{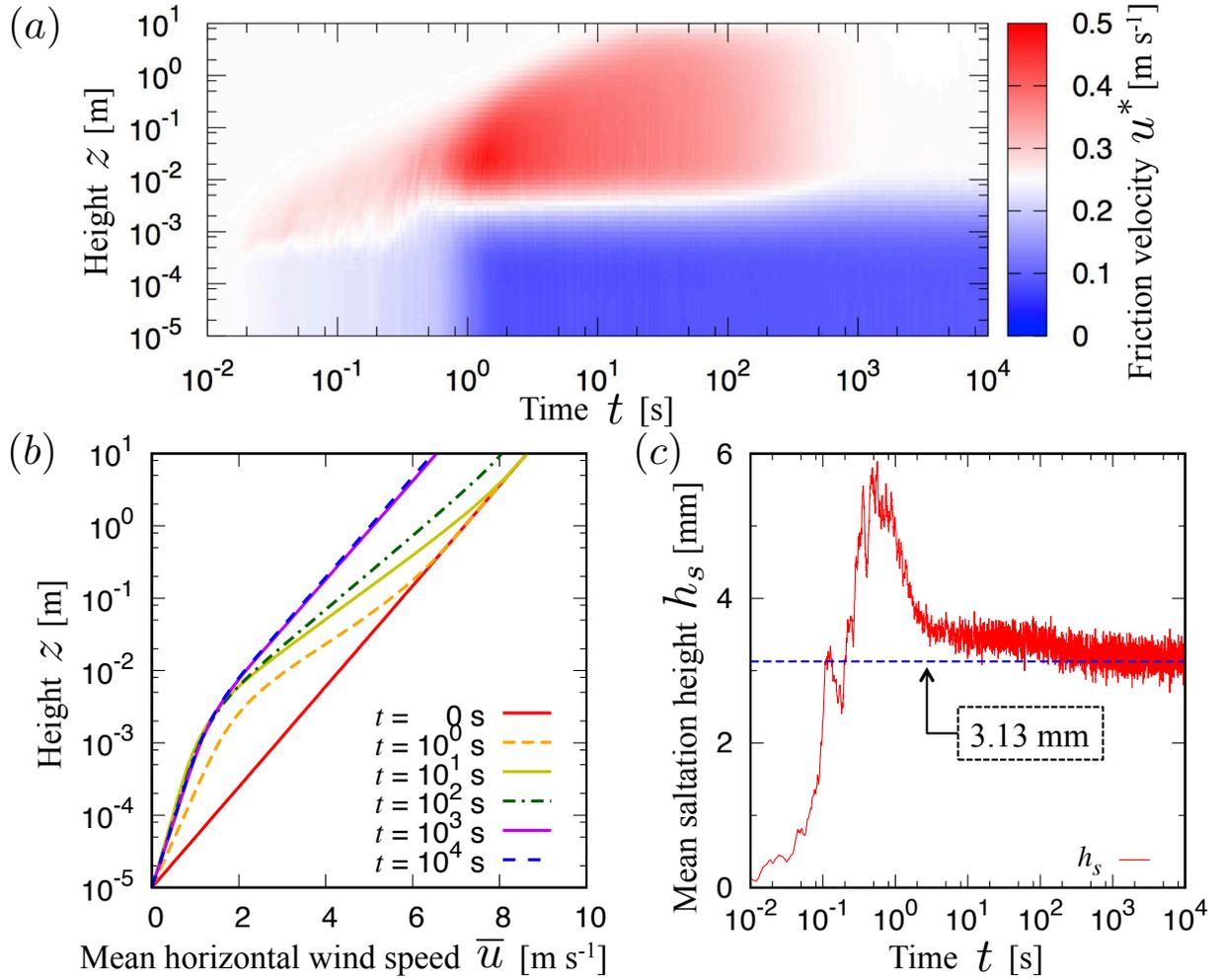}
\caption{
(Color online)
Spatiotemporal structures of transport at $u^*(H) = 0.25 \hspace{0.5 ex} \mathrm{m \hspace{0.5 ex} s^{-1}}$:
(a) friction velocity $u^*(z)$, (b) mean horizontal wind speed $\overline{u}$, and (c) mean saltation height $h_s$ defined by eq.~(\ref{eq:mean-saltation-height}).
(a) The white area denotes the initial friction velocity $u^*(t=0,z) = 0.25 \hspace{0.5 ex} \mathrm{m \hspace{0.5 ex} s^{-1}}$.
(c) The dashed line ($3.13 \hspace{0.5 ex} \mathrm{mm}$) is the time average of $h_s$ for 1 s after $t = 10^4 \hspace{0.5 ex} \mathrm{s}$.
}
\label{fig:5}
\end{figure}
%

Figure~\ref{fig:5}(a) shows the spatiotemporal structure of friction velocity $u^*$ as a color map,
where the white color denotes the initial value $u^* = 0.25 \hspace{0.5 ex} \mathrm{m \hspace{0.5 ex} s^{-1}}$.
The mean wind speed profiles $\overline{u}(z)$ at $t = 0, 1, 10, 10^2, 10^3$, and $10^4 \hspace{0.5 ex} \mathrm{s}$ are also shown in Fig.~\ref{fig:5}(b).
Because the top friction velocity $u^*(H)$ is fixed throughout the simulation,
$\overline{u} (H)$ can vary with time.
Additionally, the gradient of $\overline{u}$ plotted on the logarithmic scale for height roughly indicates the vertical profile of $u^*(z)$ because of the relationship of eq.~(\ref{eq:initial-condition}): $\overline{u} \propto u^* \log(z)$.
For the height of the particle,
we define $h_s$ as the ensemble average of particle height:
\begin{eqnarray}
h_s
=
\frac{1}{N_s} \sum_{z_i \le 10 \hspace{0.5 ex} \mathrm{cm}}^{N_s} z_i,
\label{eq:mean-saltation-height}
\end{eqnarray}
where $N_s$ is the number of particles below $10 \hspace{0.5 ex} \mathrm{cm}$ and $z_i$ is the height of the $i$th particle.
According to Nemoto et al.~\cite{nemoto2004numerical},
the boundary between saltation and suspension is $z \approx 10 \hspace{0.5 ex} \mathrm{cm}$.
Therefore, the $h_s$ of eq.~(\ref{eq:mean-saltation-height}) seems to strongly reflect the effect of saltation,
and it is named as the mean saltation height.
Figure~\ref{fig:5}(c) shows the time evolution of $h_s$.
Using Figs.~\ref{fig:5}(a), (b), and (c),
the three phases for the state of particle transport are described as follows.

(i) Development phase ($t < 1 \hspace{0.5 ex} \mathrm{s}$):
The initial friction velocity is spatially uniform as $u^*(t = 0, z) = 0.25 \hspace{0.5 ex} \mathrm{m \hspace{0.5 ex} s^{-1}}$ because the logarithmic wind speed profile $\overline{u}(t = 0, z)$ is given in eq.~(\ref{eq:initial-condition}).
The particle entrainment becomes active after $t = 10^{-2} \hspace{0.5 ex} \mathrm{s}$ (Fig.~\ref{fig:4}(a)),
and the mean saltation height $h_s$ increases with the momentum transfer from the wind (Fig.~\ref{fig:5}(c)).
In contrast, the mean horizontal wind speed $\overline{u}$ rapidly decreases below $z \approx 10 \hspace{0.5 ex} \mathrm{cm}$,
but $\overline{u}$ does not change above $z \approx 10 \hspace{0.5 ex} \mathrm{cm}$ (Fig.~\ref{fig:5}(b)).
This change in $\overline{u}$ shows that the effect of blown particles rapidly acts on the wind near the surface.
Additionally, two different gradients of $\overline{u}$ expressed on the logarithmic scale are formed below $z \approx 10 \hspace{0.5 ex} \mathrm{cm}$,
and their transition height is approximately $z = 3 \hspace{0.5 ex} \mathrm{mm}$.
This causes the non-uniform profile of friction velocity $u^*(z)$ in this range (Fig.~\ref{fig:5}(a));
$u^*$ in the vicinity of the surface is less than the initial value of $0.25 \hspace{0.5 ex} \mathrm{m \hspace{0.5 ex} s^{-1}}$,
whereas the upper $u^*$ is greater.

(ii) Relaxation phase ($t < 10^{3} \hspace{0.5 ex} \mathrm{s}$):
The decreased friction velocity $u^*$ below $z = 3 \hspace{0.5 ex} \mathrm{mm}$ is kept almost constant,
as shown in the bottom part of Fig.~\ref{fig:5}(a);
that is, the wind speed profile $\overline{u}(z \le 3 \hspace{0.5 ex} \mathrm{mm})$ hardly changes with time except around $t = 1 \hspace{0.5 ex} \mathrm{s}$ (Fig.~\ref{fig:5}(b)).
This is explained by the time evolution of the mean saltation height $h_s$.
The significant change in $h_s$ stops just a few seconds after the beginning of particle transport,
and then $h_s$ gradually decreases (Fig.~\ref{fig:5}(c)).
The relatively small temporal change in $h_s$ causes the wind speed below $z = 3 \hspace{0.5 ex} \mathrm{mm}$ to reach equilibrium.
On the other hand, the $u^*$ increased above $z = 3 \hspace{0.5 ex} \mathrm{mm}$ takes the maximum value at $t \approx 1 \hspace{0.5 ex} \mathrm{s}$ (Fig.~\ref{fig:5}(a)),
which corresponds to the peak of the total mass flux $Q$ in Fig.~\ref{fig:4}(b).
The increase in $u^*$ reaches the top ($H = 10 \hspace{0.5 ex} \mathrm{m}$) at $t = 20 \hspace{0.5 ex} \mathrm{s}$,
following which it slowly decreases to the initial friction velocity of $0.25 \hspace{0.5 ex} \mathrm{m \hspace{0.5 ex} s^{-1}}$,
as shown in the white area of Fig.~\ref{fig:5}(a).
In the simulation, the top friction velocity is fixed as $u^*(H) = 0.25 \hspace{0.5 ex} \mathrm{m \hspace{0.5 ex} s^{-1}}$;
thus, the wind speed around the top changes to satisfy $u^* = 0.25 \hspace{0.5 ex} \mathrm{m \hspace{0.5 ex} s^{-1}}$~(Fig.~\ref{fig:5}(b)).
This effect propagates from $z = H$ to $z \approx 3 \hspace{0.5 ex} \mathrm{mm}$,
which causes the long-term decrease in $u^*$.

(iii) Equilibrium phase ($t \ge 10^{3} \mathrm{s}$):
Vertical profiles of friction velocity $u^*$ and mean wind speed $\overline{u}$ are fixed, as shown in Figs.~\ref{fig:5}(a) and (b).
The $u^* (z)$ changes from a constant value less than the initial one to $0.25 \hspace{0.5 ex} \mathrm{m \hspace{0.5 ex} s^{-1}}$ around $z = 3 \hspace{0.5 ex} \mathrm{mm}$,
in which the logarithmic profile of $\overline{u}(z)$ curves.
In this simulation, $z = 3 \hspace{0.5 ex} \mathrm{mm}$ is roughly the transition height for both friction velocity and wind speed.
It should be noted that this transition height corresponds to the time average of mean saltation height
($\overline{h_s} = 3.13 \pm 0.143 \hspace{0.5 ex} \mathrm{mm}$),
which is calculated using $h_s$ for 1 s after $t = 10^4 \hspace{0.5 ex} \mathrm{s}$ (Fig.~\ref{fig:5}(c)).
Additionally, the fluctuation in $h_s$ remains at the same level in this phase.

\subsection{Wind-strength dependence at equilibrium state}
\label{sec:3-2}

We investigate the properties of particle transport depending on the wind strength by varying the top friction velocity $u^*(H)$.
The typical temporal change in particle transport is similar to the case of $u^* (H) = 0.25 \hspace{0.5 ex} \mathrm{m \hspace{0.5 ex} s^{-1}}$ despite the difference in $u^* (H)$;
hence, transport properties in the equilibrium phase ($t > 10^3 \hspace{0.5 ex} \mathrm{s}$) are shown here.

\subsubsection{Particle transport rate}
\label{sec:3-2-1}

\begin{figure}[t]
\centering
\includegraphics[width=1.0\linewidth]{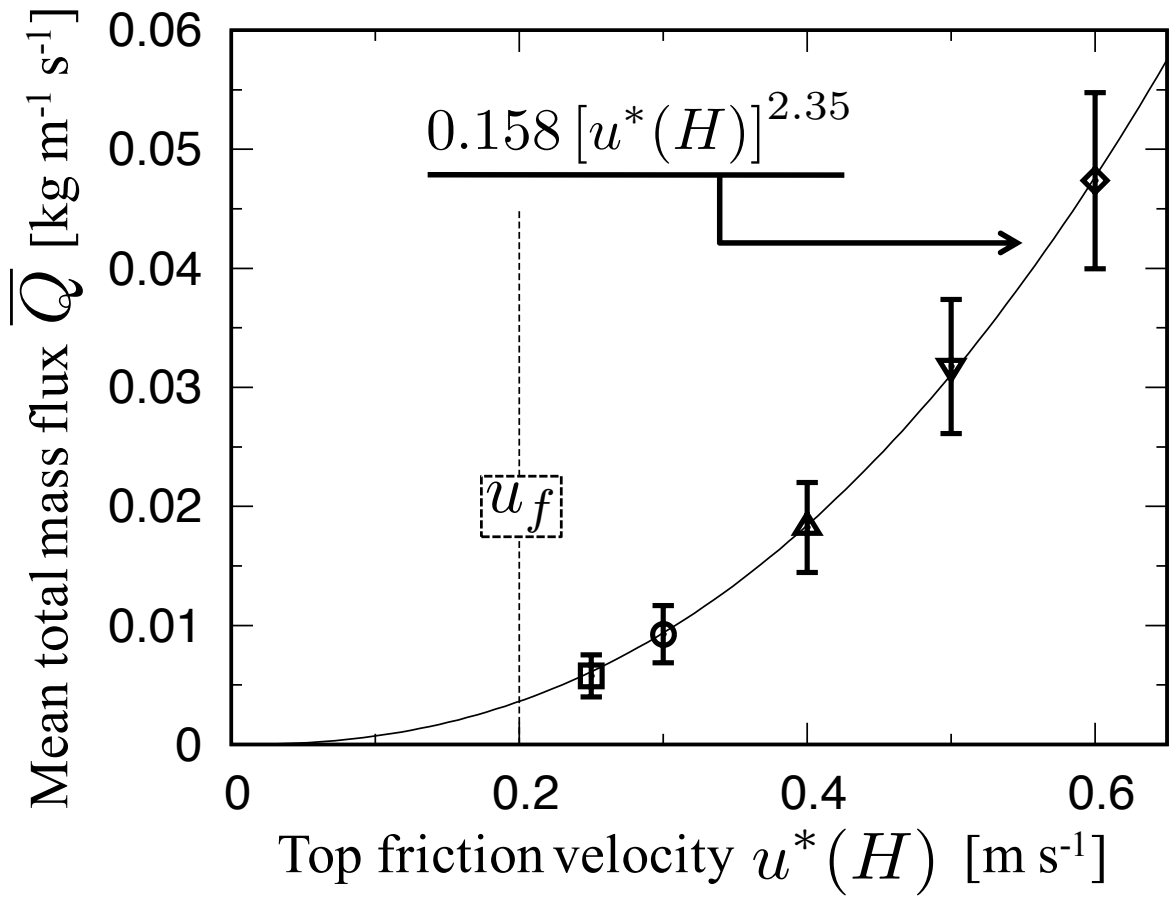}
\caption{
Time average of total mass flux $\overline{Q}$ at various top friction velocities $u^*(H)$ in equilibrium phase.
Points and error bars are calculated using the total mass flux $Q$ for 10 s after $t = 10^4 \hspace{0.5 ex} \mathrm{s}$.
The solid line is the function of $u^*(H)$ estimated using the least-squares method,
and the dashed line indicates a fluid threshold $u_f = 0.2 \mathrm{m \hspace{0.5 ex} s^{-1}}$ set in simulations.
}
\label{fig:6}
\end{figure}
%

We use the total mass flux $Q$ defined in eq.~(\ref{eq:total-mass-flux}),
but the variation of $Q$ is extremely high, as shown in Fig.~\ref{fig:4}(b).
Thus, we define $\overline{Q}$ as the time average of $Q$ for 10 s after $t = 10^4 \hspace{0.5 ex} \mathrm{s}$;
hereafter, $\overline{Q}$ is called the mean total mass flux.
Figure~\ref{fig:6} shows the relationship between top friction velocity $u^*(H)$ and $\overline{Q}$.
This relationship is well fitted by a power function of $u^*(H)$:
\begin{eqnarray}
\overline{Q}
=
0.158
\left[
u^* (H)
\right]^{2.35}.
\label{eq:friction-velocity-flux}
\end{eqnarray}
%
This power function is well known as one of the properties obtained in many previous studies~\cite{bagnold1941physics, owen1964saltation, lettau1978experimental, ungar1987steady, sugiura1998measurements, mann2000profile, nemoto2004numerical, duran2011aeolian},
but the formulation slightly differs in previous studies.
Especially, the power index of $u^*$ depends on the mode of particle dynamics: saltation and suspension.

Saltation mass fluxes are proportional to the cube of friction velocity $u^*$,
as proposed by Bagnold~\cite{bagnold1941physics}, Owen~\cite{owen1964saltation}, and Lettau et al.~\cite{lettau1978experimental}.
They assume that the speed of a saltation particle increases with $u^*$,
although more recent studies show that this assumption is not correct near the surface.
Ungar et al.~\cite{ungar1987steady} and Duran et al.~\cite{duran2011aeolian} showed that saltation mass fluxes are proportional to the square of $u^*$.
In the saltation transport with a size distribution, the power index of $u^*$ is affected by the distribution width but expected to range from 2 to 3.
On the other hand, the power index for suspension transport is generally higher than that for saltation transport,
since the increase in $u^*$ enhances the turbulence effect that drifts fine particles upward.
Indeed, Mann et al.~\cite{mann2000profile} observed suspended drifting snow in Antarctica, and they reported that $Q \propto (u^*)^{5.14}$.
This higher power index was also measured in wind tunnel experiments with polydisperse snow particles by Sugiura et al.~\cite{sugiura1998measurements} ($360 \hspace{0.5 ex} \mu \mathrm{m}$ as mean diameter): $Q \propto (u^*)^{3.96}$.
Note that Sugiura et al. measured the mass flux $q(z)$ in the saltation layer ($< 10 \mathrm{cm}$),
in which the saltated particles fluctuated by the turbulence and suspended particles are included.
In our simulations, the property of total mass flux quantitatively corresponds to the saltation transport rather than the suspension transport according to eq.~(\ref{eq:friction-velocity-flux}).

\subsubsection{Vertical profile of friction velocity}
\label{sec:3-2-2}

\begin{figure}[t]
\centering
\includegraphics[width=1.0\linewidth]{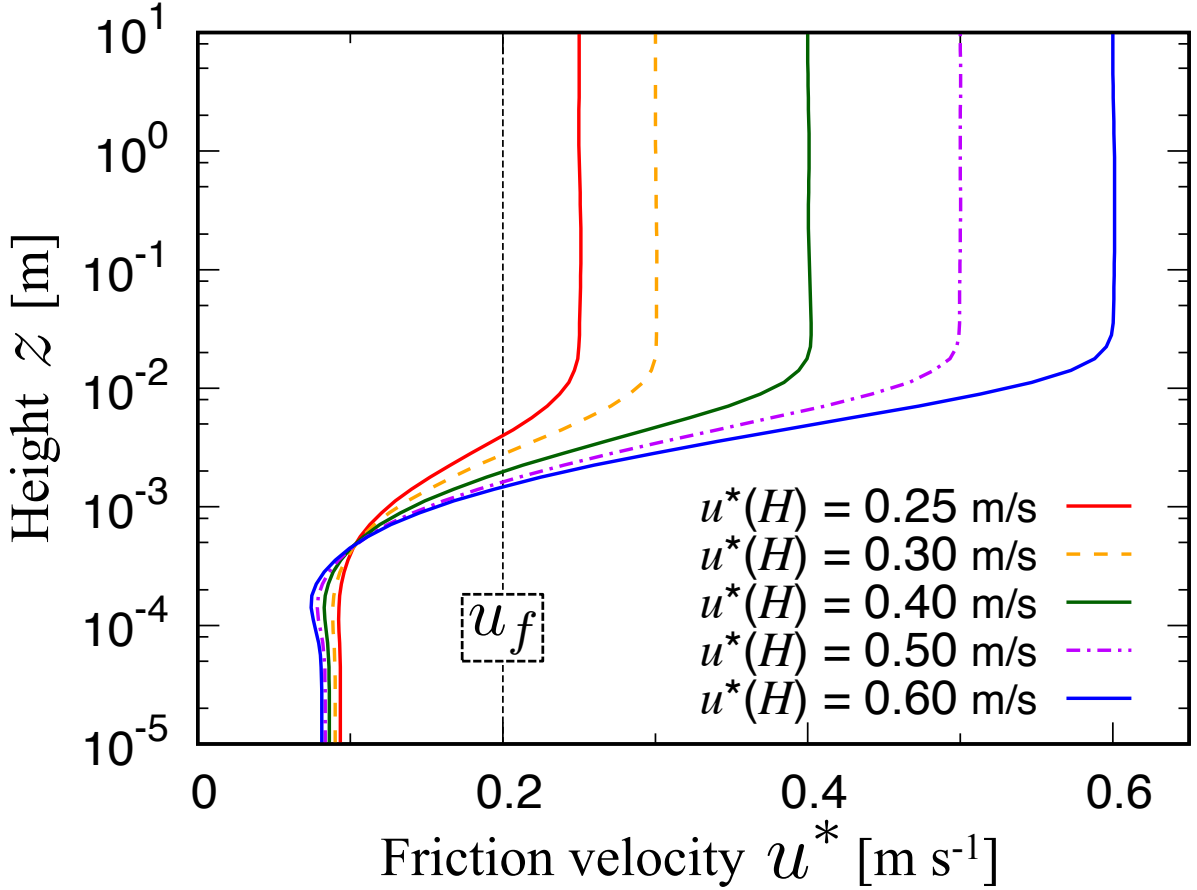}
\caption{
(Color online)
Vertical profiles of friction velocity $u^*(z)$ at various top friction velocities $u^*(H)$ in equilibrium phase ($t = 10^4 \hspace{0.5 ex} \mathrm{s}$).
The fluid threshold $u_f = 0.2 \hspace{0.5 ex} \mathrm{m \hspace{0.5 ex} s^{-1}}$ is denoted by the dashed line.
}
\label{fig:7}
\end{figure}
%

The equations of total mass flux are mostly derived on the basis of the friction velocity $u^*$ above the layer of moving particles,
but $u^*$ near the surface is spatially non-uniform as shown in Fig.~\ref{fig:5}(a).
Hence, we check the vertical profile of friction velocity $u^*(z)$ at $t = 10^4 \hspace{0.5 ex} \mathrm{s}$~(Fig.~\ref{fig:7}).
The profile of $u^*(z)$ is roughly divided into three parts according to height: $3 \hspace{0.5 ex} \mathrm{cm} < z$, $100 \hspace{0.5 ex} \mu \mathrm{m} \le z \le 3 \hspace{0.5 ex} \mathrm{cm}$, and $z < 100 \hspace{0.5 ex} \mu \mathrm{m}$.
The friction velocity above $z = 3 \hspace{0.5 ex} \mathrm{cm}$ fully reflects the top friction velocity $u^*(H)$ set in simulations,
whereas that below $z = 3\mathrm{cm}$ decreases from $u^*(H)$ because of the interaction between wind and particles.
In more detail, $u^*(100 \hspace{0.5 ex} \mu \mathrm{m} \le z \le 3 \hspace{0.5 ex} \mathrm{cm})$ decreases logarithmically with the decrease in height,
although $u^*(z < 100 \hspace{0.5 ex} \mu \mathrm{m})$ remains around $0.1 \hspace{0.5 ex} \mathrm{m \hspace{0.5 ex} s^{-1}}$ less than the fluid threshold $u_f = 0.2 \hspace{0.5 ex} \mathrm{m \hspace{0.5 ex} s^{-1}}$.
Additionally, a focus point for the friction velocity is formed at $z \approx 500 \hspace{0.5 ex} \mu \mathrm{m}$ independent of $u^*(H)$,
whereas the wall friction velocity $u_w^* \equiv u^*(z_0 = 10^{-5} \hspace{0.5 ex} \mathrm{m})$ decreases with the increase in $u^* (H)$.

\subsubsection{Relationship between wind and particle speeds}

\begin{figure}[t]
\centering
\includegraphics[width=1.0\linewidth]{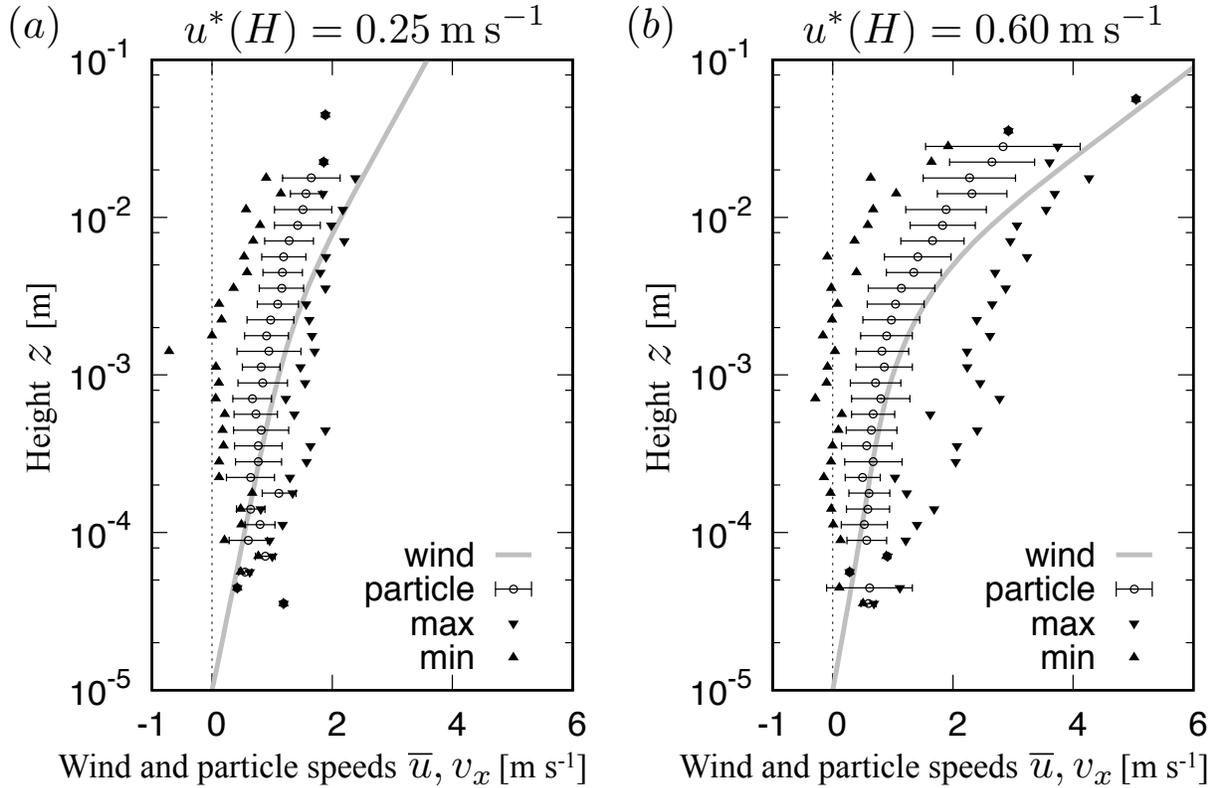}
\caption{
Vertical profiles of mean horizontal wind speed $\overline{u}(z)$ and particle speed $v_x$ in equilibrium phase ($t=10^4 \hspace{0.5 ex} \mathrm{s}$) with two types of top friction velocities $u^*(H)$:
(a) $0.25 \hspace{0.5 ex} \mathrm{m \hspace{0.5 ex} s^{-1}}$ and (b) $0.6 \hspace{0.5 ex} \mathrm{m \hspace{0.5 ex} s^{-1}}$.
The circle, error bar, and two types of triangles denote the mean, standard deviation, and maximum and minimum of $v_x$ at each height, respectively.
}
\label{fig:8}
\end{figure}
%

The total mass flux $Q$ and vertical profile of friction velocity $u^* (z)$, as shown in Figs.~\ref{fig:6} and \ref{fig:7}, respectively, are generated by particle dynamics;
thus, we focus on the wind speed and particle velocity at $t = 10^{4} \hspace{0.5 ex} \mathrm{s}$ to elucidate the relationship between them.
Figures~\ref{fig:8}(a) and (b) show the vertical profiles of the mean horizontal wind speed $\overline{u}(z)$ and horizontal particle speed $v_x$ for top friction velocities $u^*(H) = 0.25$ and $0.6 \hspace{0.5 ex} \mathrm{m \hspace{0.5 ex} s^{-1}}$, respectively.
Here, the mean, standard deviation, and maximum and minimum particle speeds are calculated for each fluid grid.
In both cases, the $\overline{u}(z)$ curve at $z \approx 3 \hspace{0.5 ex} \mathrm{mm}$ and the transition height correspond to the mean saltation height, as mentioned in \ref{sec:3-1-2} (see Figs.~\ref{fig:5}(b) and (c)).
The mean particle speed $\langle v_x \rangle$ increases with height to follow $\overline{u}(z)$ except for $z < 300 \hspace{0.5 ex} \mu \mathrm{m} = \overline{d}$ (the mean particle diameter in the granular bed),
where $\langle v_x \rangle$ exceeds $\overline{u}(z)$.
The sign inversion of speed difference between $\overline{u}(z)$ and $\langle v_x \rangle$ is related to the ascent and descent of moving particles.
We trace the typical trajectory of particles in order to confirm particle dynamics.
A particle, after collision with the surface, starts to ascend at $v_x < \overline{u}(z)$,
and the ascending particle is accelerated by the wind close to $\overline{u}(z)$.
Subsequently, the vertical movement of the particle shifts from ascent to descent,
and the $v_x$ of the descending particle exceeds $\overline{u}(z)$.
The particle is decelerated by air drag during the descent,
but it collides with the surface at a higher velocity.
The above process is repeated continually in transport.
These characteristics of ascent and descent are indicated with minimum and maximum particle speeds as triangles in Figs.~\ref{fig:8}(a) and (b), respectively.
Therefore, $\langle v_x \rangle$ near the surface is greater than $\overline{u}(z)$.

\subsubsection{Particle height distribution}

\begin{figure}[t]
\centering
\includegraphics[width=1.0\linewidth]{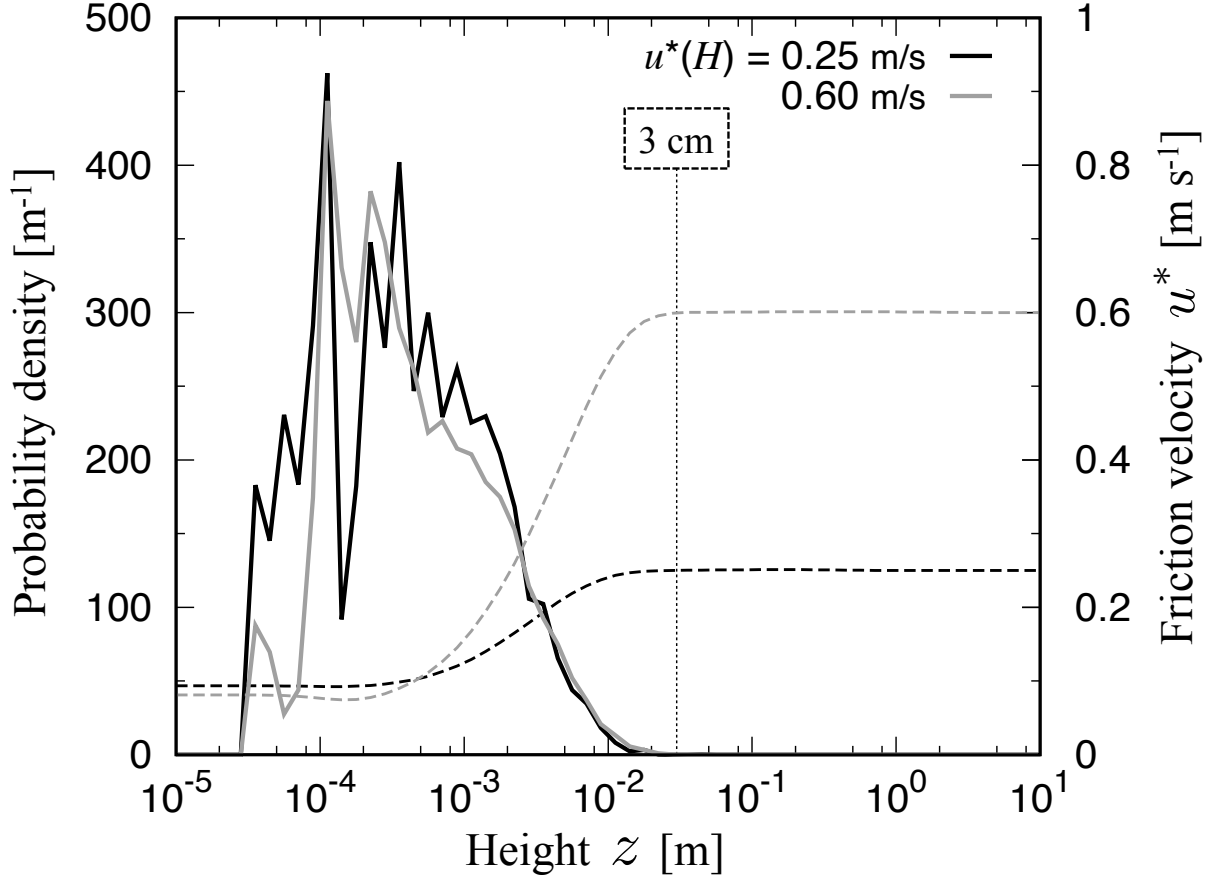}
\caption{
Probability density function of particle height and vertical profile of friction velocity $u^*(z)$ in equilibrium phase ($t = 10^4 \hspace{0.5 ex} \mathrm{s}$).
Solid and dashed lines are the probability density function and $u^*(z)$, respectively.
Line color indicates the top friction velocity $u^*(H)$: (black) $0.25 \hspace{0.5 ex} \mathrm{m \hspace{0.5 ex} s^{-1}}$ and (gray) $0.6 \hspace{0.5 ex} \mathrm{m \hspace{0.5 ex} s^{-1}}$.
}
\label{fig:9}
\end{figure}
%

The particle height seems to increase with the increase in $u^*(H)$ (Figs.~\ref{fig:8}(a) and (b)).
We calculate the probability density function of particle height to clarify the $u^*(H)$ dependency of the particle height distribution.
Figure~\ref{fig:9} shows the probability density function and the vertical profile of friction velocity $u^*(z)$ at $u^*(H) =$ $0.25 \hspace{0.5 ex} \mathrm{m \hspace{0.5 ex} s^{-1}}$ (black lines) and $0.6 \hspace{0.5 ex} \mathrm{m \hspace{0.5 ex} s^{-1}}$ (gray lines).
Both probability density functions decrease with height except for $z < 100 \hspace{0.5 ex} \mu\mathrm{m}$,
where these functions increase with height because the coordinates of particles with diameter greater than $d = 200 \hspace{0.5 ex} \mu\mathrm{m}$ cannot enter this range because of the collision with the surface.
Here, $200 \hspace{0.5 ex} \mu\mathrm{m}$ is the peak of the particle size distribution at the granular bed.
The width of the probability density function at $u^*(H) = 0.6 \hspace{0.5 ex} \mathrm{m \hspace{0.5 ex} s^{-1}}$ is nearly the same as that at $u^*(H) = 0.25 \hspace{0.5 ex} \mathrm{m \hspace{0.5 ex} s^{-1}}$,
and the tail of the function becomes zero near $z = 3 \hspace{0.5 ex} \mathrm{cm}$.
It should be noted that this height corresponds to the change point of $u^*(z)$, as mentioned in Sect.~\ref{sec:3-2-2}.
The increase in $u^*(H)$ does not affect the particle height according to the probability density functions in Fig.~\ref{fig:9}.
Despite the fact that vertical turbulent fluctuation is activated at $u^*(H) = 0.6 \hspace{0.5 ex} \mathrm{m \hspace{0.5 ex} s^{-1}}$,
there are very few particles above $z = 3 \hspace{0.5 ex} \mathrm{cm}$.
Thus, it means that particle dynamics is dominated by saltation in our simulations.
However, the maximum saltation height is almost constant ($3 \hspace{0.5 ex} \mathrm{cm}$) independent of $u^*(H)$,
and this property is inconsistent with previous studies.
For example, the wind tunnel experiment by Yang et al.~\cite{yang2007height} showed that the saltation height of sand particles monotonically increases from $u^* = 0.3 \hspace{0.5 ex} \mathrm{m \hspace{0.5 ex} s^{-1}}$ to $u^* = 0.54 \hspace{0.5 ex} \mathrm{m \hspace{0.5 ex} s^{-1}}$.
Furthermore, numerical simulations of blowing snow by Nemoto et al.~\cite{nemoto2004numerical} have reproduced the suspension of fine particles $(d < 100 \hspace{0.5 ex} \mu\mathrm{m})$ above $z = 1 \hspace{0.5 ex} \mathrm{m}$.
These differences are discussed in Sec.~\ref{sec:4}.

\subsubsection{Airborne particle diameter}
\label{sec:3-2-5}

\begin{figure}[t]
\centering
\includegraphics[width=1.0\linewidth]{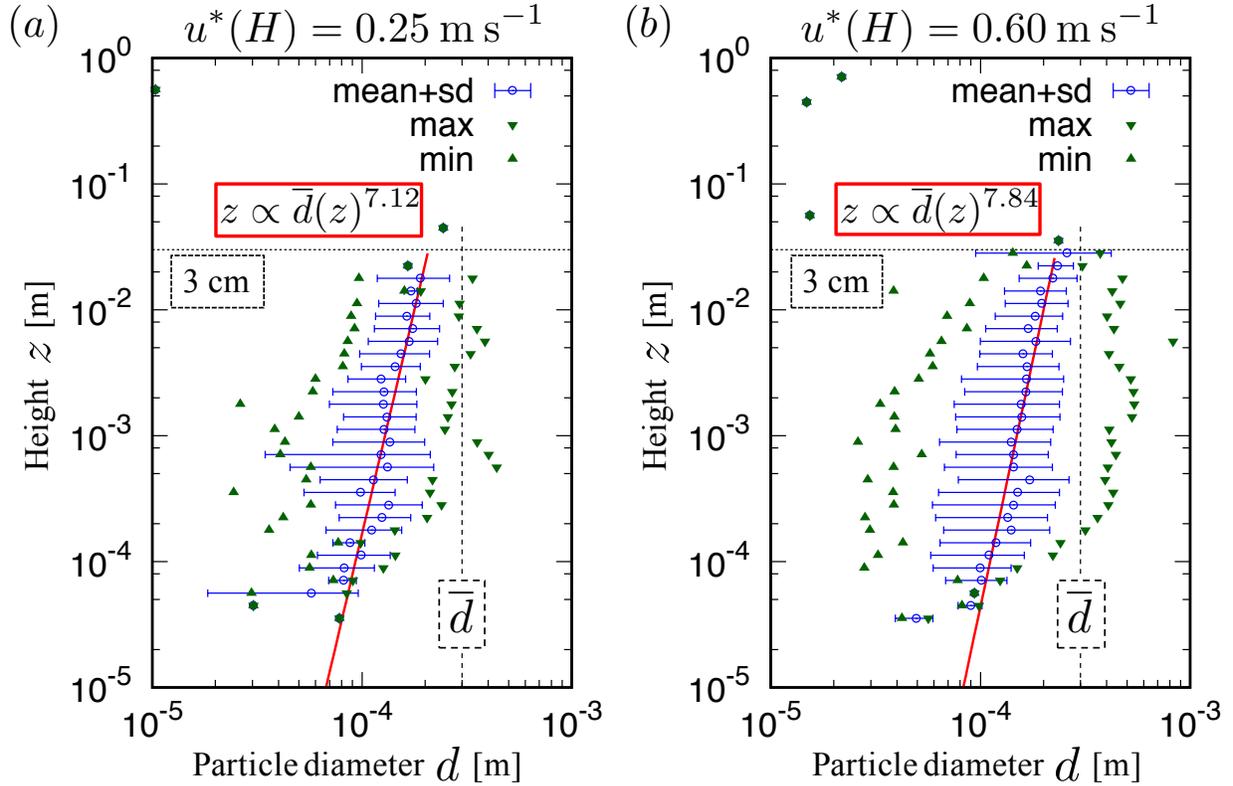}
\caption{
(Color online)
Vertical profiles of blown particle diameter in equilibrium phase ($t = 10^4 \hspace{0.5 ex} \mathrm{s}$) with two types of top friction velocities $u^*(H)$:
(a) $0.25 \hspace{0.5 ex} \mathrm{m \hspace{0.5 ex} s^{-1}}$ and (b) $0.6 \hspace{0.5 ex} \mathrm{m \hspace{0.5 ex} s^{-1}}$.
The long dashed line is the mean particle diameter ($\overline{d} = 300 \hspace{0.5 ex} \mu \mathrm{m}$) of the granular bed,
in which the gamma distribution is assumed as the particle size distribution (see Sec.~\ref{sec:2-5}),
whereas the shorter that is 3 cm height corresponding to the maximum saltation height.
The solid line below 3 cm is the power function of mean values of $d(z)$ using the least-squares method.
}
\label{fig:10}
\end{figure}
%

Particles mainly hop below $z = 3 \hspace{0.5 ex} \mathrm{cm}$ independent of the top friction velocity $u^*(H)$ (Fig.~\ref{fig:9}),
but the aeolian particle transport simulations in this study include particles from $10 \hspace{0.5 ex} \mu \mathrm{m}$ to $1 \hspace{0.5 ex} \mathrm{mm}$ in diameter.
Here, we characterize the change in particle motion depending on the diameter.
Figures~\ref{fig:10}(a) and (b) show the vertical profiles of airborne particle diameter $d$ at $u^*(H) =$ (a) $0.25 \hspace{0.5 ex} \mathrm{m \hspace{0.5 ex} s^{-1}}$ and (b) $0.6 \hspace{0.5 ex} \mathrm{m \hspace{0.5 ex} s^{-1}}$ in the equilibrium phase ($t = 10^4 \hspace{0.5 ex} \mathrm{s}$).
The mean, standard deviation, and maximum and minimum of $d$ are denoted by the circle, error bar, and two types of triangles, respectively.
In both cases, the mean particle diameter below $3 \hspace{0.5 ex} \mathrm{cm}$ increases with height: approximately 100-300 $\mu\mathrm{m}$,
and the vertical profile of mean diameter of airborne particles is roughly fitted by a power function (the solid line below 3 cm in Fig.~\ref{fig:10}):
\begin{eqnarray}
z = 10^{l}~
\overline{d}(z)^{n}
,
\end{eqnarray}
where $l$ and $n$ are fitted as (a) $l = 24.7, n = 7.12$ and (b) $l = 27.0, n = 7.84$, respectively.
On the other hand, the mean particle diameter above $3 \hspace{0.5 ex} \mathrm{cm}$ is less than that below $3 \hspace{0.5 ex} \mathrm{cm}$, since only some fine particles move up from the surface.


\section{Discussions}
\label{sec:4}

\begin{figure}[t]
\centering
\includegraphics[width=1.0\linewidth]{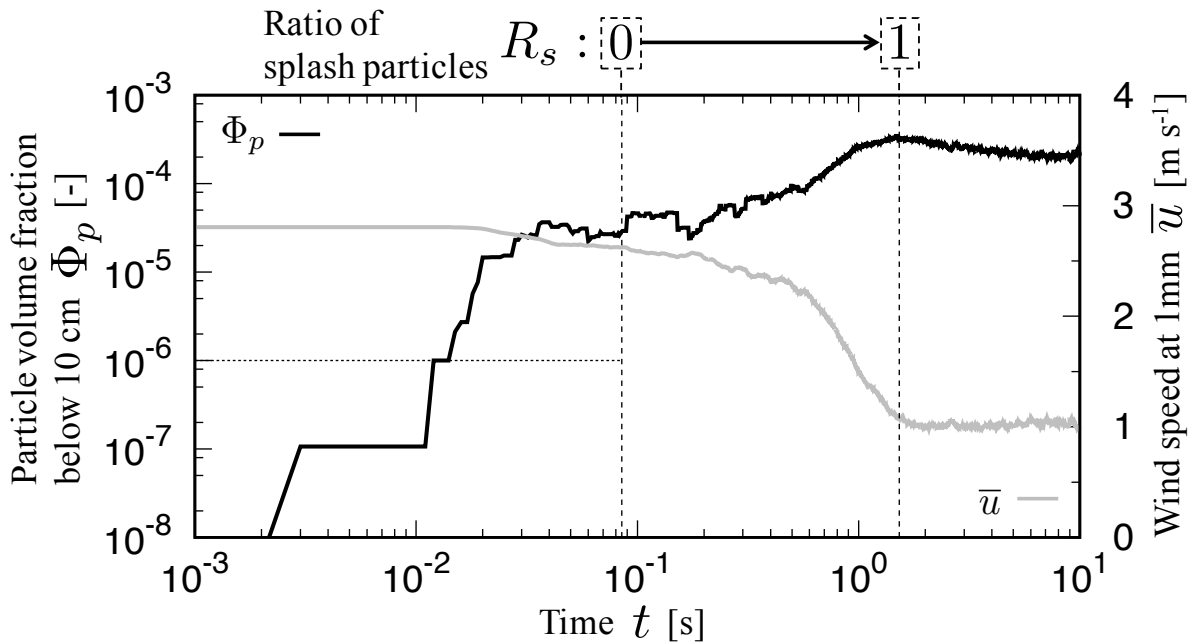}
\caption{
Time-series data of particle volume fraction below 10 cm $\Phi_p$ and mean horizontal wind speed $\overline{u}(z = 1~\mathrm{mm})$ at $u^*(H) = 0.25~\mathrm{m~s^{-1}}$.
Black and gray lines are $\Phi_p$ and $\overline{u}(z = 1~\mathrm{mm})$, respectively.
Long dashed lines indicates $R_s =$~0 and 1, respectively;
$R_s$ is the ratio of splashed particles to total number.
}
\label{fig:11}
\end{figure}

In the development phase ($t < 1$~s) of the particle transport,
the entrainment of particles shifts from the wind to splash processes with time (Fig.~\ref{fig:4}(a)).
Since the aerodynamical entrainment of our model is expressed as a increasing function of wall friction velocity $u_w^*$ in eq.~(\ref{eq:number-particles}),
the decrease in the number of particles entrained by the wind means the reduction in $u_w^*$.
Figure~\ref{fig:5}(a) actually shows that the friction velocity near the surface gradually decreases and then it is lower than the fluid threshold $u_f$.
Namely, the shift of particle entrainment is caused by the decrease in the wind speed due to blown particles (Fig.~\ref{fig:5}(b)).
To quantify the effect of particles on the wind speed,
we measure the particle volume fraction $\Phi_p$ in the saltation layer:
\begin{eqnarray}
\Phi_p
=
\frac{1}{V_f}
\sum_{z_i \le 10~\mathrm{cm}}^{N_s} V_i,
\label{eq:volume-fraction}
\end{eqnarray}
where $V_f$ and $V_i$ are the volume of fluid and $i$th particle below 10 cm, respectively.
Figure~\ref{fig:11} shows the time-series data of $\Phi_p$ and mean horizontal wind speed $\overline{u}$ at $z = 1$~mm.
The ratio of splashed particles to total number $R_s$ is also denoted by two long dashed lines (i.e., $R_s =$~0 and 1).
It is known that particles affect the air flow in approximately $\Phi_p > 10^{-6}$;
in fact, $\overline{u}$ decreases by the increase in $\Phi_p$.
In particular, the drastic increase in $\Phi_p$ leads to the rapid decrease in $\overline{u}$
while $R_s$ changes from 0 to 1 ($t \approx$~0.1-1~s).
This means the active momentum transfer from the wind to particles;
furthermore, the time scale of development phase ($\approx$~1 s) is determined by 
the activation of wind-particle interaction through the chain of splash processes.
%

After the development phase,
the particle transport state changes into the relaxation phase ($t < 10^3$~s) according to the decrease in the total mass flux $Q$ (Fig.~\ref{fig:4}(b)).
The wind speed $\overline{u}$ (or friction velocity $u^*$) below 3 mm is nearly unchanged since $t =$~10 s,
whereas that above 3 mm gradually decreases with time because of $u^*$ fixed at the top (Fig.~\ref{fig:5}(a) and (b)).
The mean height of saltation particles is also lower to reflect this decrease in $\overline{u}$ as shown in Fig.~\ref{fig:5}(c).
Hence, the time scale of relaxation phase $t = 10^3$~s seems to be caused by only the decrease in $\overline{u}(z > 3~\mathrm{mm})$.
To evaluate this relaxation time,
we discuss it from the time evolution of $\overline{u}$ expressed by eq.~(\ref{eq:mean-flow}).
Firstly, the range of 3 mm~$\le z \le$~10 m is treated as a single fluid grid.
Secondly, we assume that particles does not affect the wind and the gradient of fluid shear stress $\partial \tau / \partial z$ is spatially uniform in this fluid grid.
In fact, we confirm that $\tau$ decreases approximately linearly with height since $t = 10^2$~s.
By conducting the first-order accurate discretization of eq.~(\ref{eq:mean-flow}) based on above assumptions,
the relaxation time to reach the equilibrium $T_{r}$ is roughly expressed as:
\begin{eqnarray}
T_{r}
=
\rho_f
|\Delta \overline{u}|
\frac{\Delta z}{|\Delta \tau|},
\label{eq:relaxation-time}
\end{eqnarray}
where $\rho_f$ is the fluid density,
$\Delta \overline{u}$ is the wind speed difference from the equilibrium value,
$\Delta z$ is the height of fluid grid,
and $\Delta \tau$ is the fluid shear stress difference between the bottom and top of fluid grid.
We substitute numerical values at $t = 10^2$~s into eq.~(\ref{eq:relaxation-time}),
and $T_r \approx 266$~s is obtained:
$|\Delta \overline{u}| \approx 1.44$~$\mathrm{m \hspace{0.5 ex} s^{-1}}$ ($z = 5$~m),
$\Delta z \approx 10$~m,
$|\Delta \tau| \approx 0.065$~$\mathrm{kg~m^{-1}~s^{-2}}$.
The actual relaxation time ($10^3$~s) is longer than this estimation,
although it appears because of the convergence of $|\Delta \tau|$ to 0.

The relaxation time in our simulations $t = 10^3$~s is extremely longer than that of wind tunnel experiments~\cite{yang2007height, walter2014experimental, nishimura2014snow} and previous numerical simulations~\cite{shao1999numerical, nemoto2004numerical, huang20153}.
In the previous studies,
the wind tunnels have a total length greater than $10 \hspace{0.5 ex} \mathrm{m}$, and the general measurement of mass flux $q(z)$ is conducted approximately $10 \hspace{0.5 ex} \mathrm{m}$ from the inlet or the particle supply point.
The particle transport is assumed to reach the quasi-equilibrium or equilibrium state at the measuring point.
When this occurs and the mean horizontal particle velocity is 1-2 $\mathrm{m \hspace{0.5 ex} s^{-1}}$,
the relaxation time is roughly estimated as 5-10 $\mathrm{s}$.
The relaxation time, of the order of 10 $\mathrm{s}$, has been reproduced by previous numerical simulations,
where a constant wind speed is set at the top boundary.
The boundary condition reflects the free-stream wind velocity of the wind tunnel,
but the height of free-stream is approximately the center height of wind tunnel: 50 cm,
which is much lower than that of natural fields.
As the wind speed in natural fields is not constant but variable from hour to hour,
the particle transport is reviewed according to the friction velocity, expressing the logarithmic profile of the wind speed~\cite{mann2000profile, nishimura2005blowing}.
In our simulations, the wind speed at the top is variable with time since the friction velocity is fixed at the top;
thus, the boundary condition is better than that of previous simulations in the elucidation of the transport property at the constant friction velocity.
In addition, the wind speed profile below $z = 1 \hspace{0.5 ex} \mathrm{m}$ varies during $t = 10 \hspace{0.5 ex} \mathrm{s}$,
whereas the wind speed at the top does not change during the time (Fig.~\ref{fig:5}(b)),
which quantitatively corresponds to the boundary condition fixed the wind speed at the top.

The time-averaged total mass flux $\overline{Q}$ is well expressed with the power function of top friction velocity $u^*(H)$ in the equilibrium state, as is generally well known~(see Sec.~\ref{sec:3-2-1}).
It should be noted that the power index of $u^*(H)$ strongly depends on the saltation and suspension of particles.
Indeed, the transport consisting of pure saltation shows a power index of 2-3,
whereas the transport including both saltation and suspension exhibits a power index greater than 3.
Although we consider a vertical turbulent effect acting on particles,
the power index is obtained as 2.35.
This property is caused by the lack of suspension particles, as shown in Fig.~\ref{fig:9}.
From the above fact, we can say that most particles are transported by the saltation in our simulations,
although the turbulent effect disrupts the saltation trajectory.

The saltation height shows a weak response to $u^*(H)$ (or wind speed):
the maximum saltation height is approximately $3 \hspace{0.5 ex} \mathrm{cm}$ (Fig.~\ref{fig:9}).
The dynamics of saltation are mostly determined by splash functions applied in our model (see Sec.~\ref{sec:2-4}).
In particular, the vertical restitution coefficient $e_v$ is directly related to the particle height,
since the vertical ejected velocity $v_{ez}$ is calculated as $e_v |v_{iz}|$,
where $v_{iz}$ is the vertical incident velocity.
Figure~\ref{fig:2}(c) shows the effect of both incident speed and angle on the probability density function of $e_v$.
As the increase in incident speed shifts the peak of the distribution to zero,
the vertical ejected velocity is not increased drastically.
This characteristic of $e_v$ is associated with the low saltation height at relatively high wind speeds.
According to wind tunnel experiments by Yang et al.~\cite{yang2007height}, the saltation height monotonically increases with increase in the friction velocity.
Therefore, we should improve splash functions measured by Sugiura et al.~\cite{sugiura2000wind} in friction velocities greater than the measurement range: that is, the lower incident angle and higher incident speed.

\begin{figure}[t]
\centering
\includegraphics[width=1.0\linewidth]{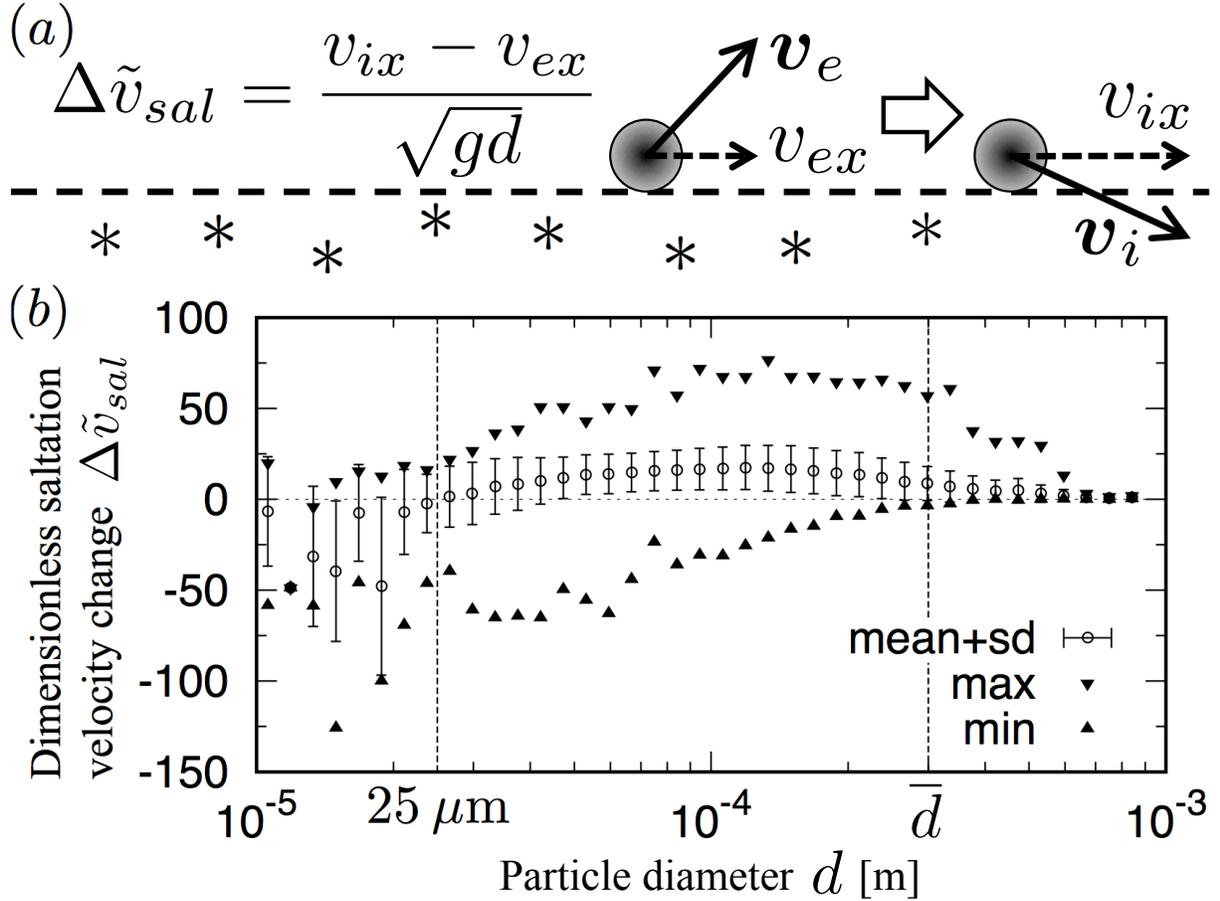}
\caption{
Acceleration and deceleration of saltation particle at $u^*(H) = 0.6~\mathrm{m \hspace{0.5 ex} s^{-1}}$.
(a) Definition of dimensionless saltation velocity change $\Delta \tilde{v}_{sal}$ at beginning and end of a single saltation;
$v_{ex}$ and $v_{ix}$ are horizontal ejection and incident velocities, respectively.
(b) $\Delta \tilde{v}_{sal}$ for various particle diameters;
all incident particles for 1~$\mathrm{s}$ after $t = 10^4~\mathrm{s}$ are used for the calculation of $\Delta \tilde{v}_{sal}$.
}
\label{fig:12}
\end{figure}

The lack of suspension also occurs at higher friction velocity $u^*(H) = 0.6 \hspace{0.5 ex} \mathrm{m \hspace{0.5ex} s^{-1}}$ because fine particles cannot move up as snow in Fig.~\ref{fig:10}(b).
We quantify particle acceleration and deceleration due to the wind in order to reveal the detailed diameter dependency of saltation particles.
As a simple indicator, the dimensionless saltation velocity change $\Delta \tilde{v}_{sal}$ is proposed (Fig.~\ref{fig:12}(a)):
\begin{eqnarray}
\Delta \tilde{v}_{sal}
=
\frac{v_{ix} - v_{ex}}{\sqrt{g d}},
\label{eq:dimensionless-velocity}
\end{eqnarray}
where $g$ is the gravitational acceleration,
$d$ is the particle diameter,
and $v_{ex}$ and $v_{ix}$ are horizontal ejected and incident velocities for the single-particle saltation, respectively.
Positive and negative values of $\Delta \tilde{v}_{sal}$ indicate acceleration and deceleration through the air drag.
Figure~\ref{fig:12}(b) shows $\Delta \tilde{v}_{sal}$ for various particle diameters at the top friction velocity $u^*(H) = 0.6 \hspace{0.5 ex} \mathrm{m \hspace{0.5ex} s^{-1}}$,
and $\Delta \tilde{v}_{sal}$ values are calculated using all incident particles for $1 \hspace{0.5 ex} \mathrm{s}$ after $t = 10^4 \hspace{0.5 ex} \mathrm{s}$.
Here, the mean, standard deviation, maximum, and minimum of $\Delta \tilde{v}_{sal}$ are denoted by the circle, error bar, and two types of triangles, respectively.
Fine particles smaller than $d = 25 \hspace{0.5 ex} \mu\mathrm{m}$ show a negative $\Delta \tilde{v}_{sal}$,
which implies deceleration during the migration.
In addition, the turbulent intensity $\sigma_w$ acting on the particle is too weak near the surface,
since $\sigma_w$ defined in eq.~(\ref{eq:time-intensity}) is proportional to the friction velocity $u^*$,
and $u^*$ at the vicinity of the surface is much less than the top friction velocity $u^*(H)$ (Fig.~\ref{fig:7}).
Both the deceleration of fine particles and decrease in $\sigma_w$ prevent fine particles from ascending.


%
%

Furthermore, particles with $d > 25 \hspace{0.5 ex} \mu\mathrm{m}$ are mainly accelerated by the wind,
as their dimensionless saltation velocity change $\Delta \tilde{v}_{sal}$ is positive in Fig.~\ref{fig:12}(b).
The particles ranging from $100 \hspace{0.5 ex} \mu\mathrm{m}$ to $300 \hspace{0.5 ex} \mu\mathrm{m}$ ($= \overline{d}$: mean particle diameter of granular bed) shows the peak of mean $\Delta \tilde{v}_{sal}$,
which indicates the more effective particle acceleration due to the wind.
As a result of the effective acceleration,
the space is occupied by particles with approximately 100-300 $\mu\mathrm{m}$ in diameter (Fig.~\ref{fig:10}(b)).
The airborne particle diameter increases with height as shown in Figs~\ref{fig:10}(a) and (b),
but this property is inconsistent with some previous studies~\cite{gromke2014snow, nishimura2005blowing}.
The wind tunnel experiment by Gromke et al.~\cite{gromke2014snow} shows that the mean snow particle diameter is fairly constant with height in the saltation layer,
whereas Nishimura et al.~\cite{nishimura2005blowing} found from a field observation of blowing snow in Antarctica that the particle diameter distribution can be approximated by a gamma distribution, which moves to smaller diameters with height.
These results show that the mean particle diameter decreases from the saltation layer to the suspension layer.
In both studies, the fine particle exhibits the suspension,
although that of our simulations does not drift up from the surface.
Hence, the lack of suspension is related to the increase in diameter with height.
Note that both measurements of diameter were taken above $z \approx 1 \hspace{0.5 ex} \mathrm{cm}$;
therefore, the particle diameter at $z < 1 \hspace{0.5 ex} \mathrm{cm}$ is still not known well.

The vertical profile of friction velocity $u^*(z)$ at the equilibrium state is divided into three ranges according to height (Fig.\ref{fig:7}):
$z < 100 \hspace{0.5 ex} \mu\mathrm{m}$, $100 \hspace{0.5 ex} \mu\mathrm{m} \le z \le 3 \hspace{0.5 ex} \mathrm{cm}$, and $3 \hspace{0.5 ex} \mathrm{cm} < z$,
where $100 \hspace{0.5 ex} \mu\mathrm{m}$ is the peak of the probability density function of particle height,
and $3 \hspace{0.5 ex} \mathrm{cm}$ well corresponds to the maximum saltation height (Fig.~\ref{fig:9}).
It should be noted that $u^*(100 \hspace{0.5 ex} \mu\mathrm{m} \le z \le 3 \hspace{0.5 ex} \mathrm{cm})$ decreases logarithmically with decrease in height but $u^*(z < 100 \hspace{0.5 ex} \mu\mathrm{m})$ remains roughly constant:
0.08-0.1 $\mathrm{m \hspace{0.5 ex} s^{-1}}$.
That is, the wall friction velocity $u_w^*$ acting on the surface is always less than the fluid threshold $u_f = 0.2 \hspace{0.5 ex} \mathrm{m \hspace{0.5 ex} s^{-1}}$ set in simulations.
According to Owen~\cite{owen1964saltation},
$u_w^*$ is equal to the impact threshold $u_i$ at the equilibrium state of particle transport,
which is well known as Owen's hypothesis.
Here, $u_i$ denotes the minimum friction velocity required to maintain particle transport.
However, recent wind tunnel experiments~\cite{walter2014experimental} and numerical simulations~\cite{kok2012physics} have found that Owen's hypothesis does not hold true in some cases,
although $u_w^* < u_f$ is satisfied in all cases.
The behavior of $u_w^*$ in our simulations is qualitatively consistent with previous studies~\cite{owen1964saltation, walter2014experimental, kok2012physics},
but we should study $u_w^*$ in detail as future work.

The most important point of discussion is the comparison between our simulations and results of the random-flight model proposed by Nemoto et al.~\cite{nemoto2004numerical},
since their model is the original considered in this study.
Key differences from their model are noted in the following three aspects:
(i) the boundary condition for the wind at the top, (ii) the calculation method of mean horizontal wind speed $\overline{u}$, and (iii) the method used for splash functions.
(i)
During simulations, Nemoto et al. fixed the wind speed $\overline{u}$ at $z = 20 \hspace{0.5 ex} \mathrm{m}$ as the top,
whereas we fix the friction velocity $u^*$ (or fluid shear stress) at $z = 10 \hspace{0.5 ex} \mathrm{m}$ as the top.
As discussed at the beginning of this section,
this difference in the boundary condition at the top affects the relaxation time of particle transport to reach the equilibrium state;
that is, the relaxation time of our simulations ($\approx 10^{3} \hspace{0.5 ex} \mathrm{s}$) is much longer than that of their simulations ($\approx 10 \hspace{0.5 ex} \mathrm{s}$).
It should be noted that our boundary condition quantitatively consists with their boundary condition only on a time scale of 10 s,
because the wind speed at the top does not change on the time scale, as shown in Fig.~\ref{fig:5}(b).
In addition, since the friction velocity of wind profile at the equilibrium state is equal to that fixed at the top,
our boundary condition is better for the transport property under the accurate friction velocity.
(ii)
The horizontal uniform flow of wind speed is assumed in both the simulations, but Nemoto et al. also assumed the wind profile to be steady.
This implies that the wind speed immediately changes with the drag force due to drifting particles.
They showed that the wind speed at the equilibrium state is slower than the initial logarithmic wind profile,
which is consistent with our results (Fig.~\ref{fig:5}(b)).
On the other hand, they reported that the wall friction velocity $u_w^*$ at the equilibrium state is higher than the fluid threshold $u_f$ set in simulations, although $u_w^*$ decreases with time.
For the equilibrium property of $u_w^*$, we obtain $u_w^* < u_f$, which is opposite to Nemoto et al.'s results.
The reason is that the momentum exchange between the wind and particles might be underestimated near the surface in their simulations,
since they do not calculate the wind speed at $z < 600 \hspace{0.5 ex} \mu\mathrm{m}$.
Namely, our result ($u_w^* < u_f$) is obtained by calculating the wind-particle interaction at the vicinity of the surface.
In fact, numerical simulations by Kok et al.~\cite{kok2009comprehensive} in the range $10^{-5} \hspace{0.5 ex} \mathrm{m} \le z \le 10 \hspace{0.5 ex} \mathrm{m}$ showed that $u_w^* < u_f$ using the change in horizontal wind speed calculated by the force balance.
Therefore, the above comparison suggests that we are unable to ignore the calculations of wind speed and particle trajectory at the vicinity of the surface
because the wall friction velocity $u_w^*$ determines the aerodynamical entrainment,
which is one of the physical sub-processes in aeolian particle transport.
(iii)
Splash processes of both models are based on splash functions observed in wind tunnel experiments by Sugiura et al.~\cite{sugiura2000wind} (see Sec.~\ref{sec:2-4}).
Distributions of horizontal and vertical restitution coefficient in eq.~(\ref{eq:splash-functions}) are obtained by measuring the rebound of incident particles;
hence, we apply these distributions to the rebound.
On the other hand, Nemoto et al.~\cite{nemoto2004numerical} calculate the rebound by formulas proposed by McEwan et al.\cite{mcewan1993adaptation},
in which the rebound restitution coefficient and rebound angle are functions of incident angle and uniform random numbers.
In both models, the dynamics of splash particles (i.e., new particles ejected from the bed) is simulated by splash functions measured by Sugiura et al.~\cite{sugiura2000wind}.
Therefore, the calculation of rebound particles is difference from Nemoto et al.'s model,
and our model correctly reproduces results of wind tunnel experiments by Sugiura et al~\cite{sugiura2000wind} than their model.
As reported by Nemoto et al.~\cite{nemoto2004numerical},
coarse particles with diameter greater than $d = 100 \hspace{0.5 ex} \mu\mathrm{m}$ show saltation below $z = 10 \hspace{0.5 ex} \mathrm{cm}$,
whereas fine particles with diameter less than $d = 100 \hspace{0.5 ex} \mu\mathrm{m}$ exhibit suspension above $z = 1 \hspace{0.5 ex} \mathrm{m}$.
However, these particle heights are not confirmed by the vertical profiles of particle diameter in our simulations, as shown in Figs.~\ref{fig:10}(a) and (b).
This is related to the difference in formulation of the dynamics of rebound particles,
although minor modifications are conducted from Nemoto et al.'s model.
Here, splash functions by Sugiura et al.~\cite{sugiura2000wind} were measured at low friction velocities close to the fluid threshold $u_f$;
thus, it is unclear whether their splash functions reproduce the particle dynamics at higher friction velocities.
Their splash functions should be improved for higher friction velocities
because our numerical simulations show the maximum saltation height of 3 cm and the lack of suspension at the highest friction velocity ($0.6 \hspace{0.5 ex} \mathrm{m \hspace{0.5 ex} s^{-1}}$), as shown in Fig.~\ref{fig:10}(b).
Moreover, we should divide the particle dynamics into rebound and splash,
and the diameter dependence of splash processes shown by single splash experiments~\cite{beladjine2007collision, ammi2009three} should also be taken into account.

\section{Conclusions}

In this study, we calculated the aeolian particle transport on a flat surface based on the random-flight model~\cite{nemoto2004numerical} of blowing snow
in order to elucidate the spatiotemporal structure in the transport from the vicinity of the surface ($10^{-5} \hspace{0.5 ex} \mathrm{m}$) to $10 \hspace{0.5 ex} \mathrm{m}$ in height.
The splash process, one of the physical sub-processes in the model, is expressed by splash functions measured in wind tunnel experiments~\cite{sugiura2000wind} with snow particles.
This method is suitable for the simulation of prolonged transport,
since the complicated collision process between particles in the granular bed is simplified.
As the boundary condition,
we fixed the friction velocity at the top, where the wind speed can vary with time.
This boundary condition is superior to that of constant wind speed at the top in the elucidation of transport property under a constant friction velocity.

Our numerical results are summarized as follows.
(i)
The temporal change in typical particle transport is classified into three phases according to the particle transport rate:
development ($t < 1 \hspace{0.5 ex} \mathrm{s}$), relaxation ($t < 10^3 \hspace{0.5 ex} \mathrm{s}$), and equilibrium ($t \ge 10^3 \hspace{0.5 ex} \mathrm{s}$).
These phases are formed by wind weakening in two steps:
rapid response below $z \approx 10 \hspace{0.5 ex} \mathrm{cm}$ and gradual response above.
(ii)
The particle transport rate at the equilibrium state is well expressed as a power function of the fixed top friction velocity,
which is a well-known property in aeolian particle transport.
We obtain a power index of $2.35$,
which indicates that particles are mostly transported by saltation.
(iii)
The friction velocity at the equilibrium state decreases from the top friction velocity below the maximum saltation height ($\approx 3 \hspace{0.5 ex} \mathrm{cm}$).
In particular, the friction velocity at $z < 100 \hspace{0.5 ex} \mu\mathrm{m}$ remains roughly constant and less than the fluid threshold set in simulations.
(iv)
The mean particle speed at $z \ge 300 \hspace{0.5 ex} \mu\mathrm{m}$ ($=$ mean particle diameter of the granular bed) is less than the wind speed,
whereas that at $z < 300 \hspace{0.5 ex} \mu\mathrm{m}$ exceeds the wind speed because of descending particles.
(v)
The airborne particle diameter increases with height in the saltation layer ($z < 3 \hspace{0.5 ex} \mathrm{cm}$),
where the relationship between mean diameter and height is well expressed as a power function.
Note that the lack of fine particles is caused by two factors:
the decrease in velocity during saltation,
and the decrease in turbulent intensity due to the lower friction velocity near the surface.
%

Finally, splash functions used in our model were measured in wind tunnel experiments by Sugiura et al.~\cite{sugiura2000wind},
where the friction velocity of wind profile closes to the fluid threshold.
Although their experimental condition is the low wind speed,
we studied the property of aeolian particle transport by utilizing splash functions to relatively high wind speed.
Splash processes calculated in the model are correct at low friction velocities corresponding to their experimental condition,
but we found the crucial problem of splash functions through comparisons with previous studies (see Sec.~\ref{sec:4}).
These functions are unable to reproduce the particle dynamics at friction velocities higher than the upper limit of the experiments,
because the vertical restitution coefficient of rebound particles is underestimated.
Therefore, we suggest that the splash functions should be improved or reconstructed on the basis of more detailed experiments at higher friction velocities.



\end{document}